\documentclass[pre,showpacs,preprint,amsmath,amssymb,aps]{revtex4-1}

\usepackage[dvipdfmx]{graphicx}

\usepackage{graphicx}
\usepackage{dcolumn}
\usepackage{bm}
\usepackage{color}
\usepackage{amsmath}

\begin{document}
 
\title{
Effect of the Surface Diffusion and Evaporation of Impurities on Step Bunching  Induced by 
Impurities 
}

\author{
Masahide Sato}
\affiliation{
Information Media Center, Kanazawa University,
Kanazawa 920-1192, Japan
}%


\date{\today}


\begin{abstract}
We consider a vicinal face, where
atoms and impurities impinge and evaporate to a vapor phase,
to study how  the surface  diffusion and evaporation  of impurities affect  
 step bunching induced by impurities.
When  the lifetime of impurities on the vicinal face  $\tau_\mathrm{imp}$ 
is long and  the surface diffusion of impurities  is neglected, 
 the step  bunches induced by impurities are tight.
When $\tau_\mathrm{imp}$ decreases,
the size of the step bunches,  which means the number of steps in the bunches,
decreases but
 the separation of single steps from bunches does  not occur.
When   we take into account  fast surface diffusion of impurities, the
separation and collision between  single steps and  step bunches occur  repeatedly.
\end{abstract}

\maketitle

\section{Introduction}\label{sec:introduction}

Step bunching is well-known to be induced by impurities.
Many groups have studied the theory underlying step 
bunching~\cite{cv,Frank,Eerden1986prl57,Kandel92prl69,Kandel-w94prl49_5554,Krug-epl2002,%
Vollmer08njp053017,Ranganathan-w13pre055503,Ranganathan-w2014jcg393-35,Ltsuko-2014cgd6129,
Ltsuko-2016prl015501,%
Sleutel-ld2018cgd18_171}.
By assuming 
that the velocity  of a single step 
decreases with increasing the impurity density on a surface,
Frank~\cite{Frank}  was able to explain  why step bunching is caused by 
impurities.  
In that model, the impurity density increases  until advancing steps  refresh the surface.
The time evolution of 
 the number of steps in bunches
during the step bunching was studied using a one-dimensional
model~\cite{Eerden1986prl57,Kandel92prl69},
in which the dependence of the step velocity on 
the terrace width was assumed empirically.
From Monte Carlo simulations,
Weeks and co-workers~\cite{Kandel-w94prl49_5554,Ranganathan-w13pre055503,
Ranganathan-w2014jcg393-35} 
were able to 
study two-dimensional step motion
 and showed that mesh-like step bunch patterns are formed by  step bunching.
In their model, the effect of impurities is taken into account as 
a reduction in  the  probability
of step advancing.
Sluetel and co-workers~\cite{Ltsuko-2016prl015501} 
also studied  step bunching induced by impurities using another type 
of model.
They showed that  macrosteps formed by impurities
can advance under   high impurity conditions
in  which elementary steps cannot advance. 

Those models used in the previous studies~\cite{cv,Frank,Eerden1986prl57,Kandel92prl69,
Kandel-w94prl49_5554,Krug-epl2002,%
Vollmer08njp053017,Ranganathan-w13pre055503,Ranganathan-w2014jcg393-35,Ltsuko-2014cgd6129,
Ltsuko-2016prl015501,%
Sleutel-ld2018cgd18_171}
are simple and useful,
but  the surface diffusion field formed by adatoms
is neglected and  motions  of impurities are not taken into account concretely.
Therefore, we developed  a model in which those processes were  adopted
and   studied the step bunching induced by impurities
performing Monte Carlo simulations~\cite{msato-jpsj2017,ms2018pre}.
We suppose that both atoms and impurities impinge from a vapor phase to a vicinal face 
with  impingement rates $F$ and $F_\mathrm{imp}$, respectively.
We  assume that impurities  are contained in materials with a constant ratio.
Under  that assumption, the ratio of $F_\mathrm{imp}$ to $F$ should be kept constant  if $F$ changes.
Therefore, we kept $F_\mathrm{imp}/F$  constant  in our simulations.
 In our previous studies~\cite{msato-jpsj2017,ms2018pre},
we neglected the surface diffusion of impurities  and 
studied how step bunching induced by impurities depends  on $F$. 
When neither  impurities   nor  adatoms evaporate from a vicinal face~\cite{msato-jpsj2017},
step bunching is caused by small $F$.
The impurity density on the surface $\sigma_\mathrm{imp}$
increases  when step bunches are formed.
We also studied  step bunching in two other systems:
one is the system in which only impurities  evaporate,
and the other is the system in which both atoms and impurities evaporate~\cite{ms2018pre}.
When only impurities evaporate,
the step bunching is induced by small $F $,
which is the same as that 
in the system where neither impurities nor  adatoms evaporate~\cite{msato-jpsj2017}.
When large bunches are formed, 
$\sigma_\mathrm{imp}$  increases
but 
the density of impurities incorporated in solid 
$\rho_\mathrm{imp}$ decreases.
When both impurities and adatoms evaporate,
$\sigma_\mathrm{imp}$ and $\rho_\mathrm{imp}$ increase because of  step bunching.

Separation of single steps from bunches did not occur
 in our previous studies~\cite{msato-jpsj2017,ms2018pre}.
We wondered why   the  separation of steps,
which is observed  during  step bunching in 
other systems~\cite{sato-u-prb11172_51_1995,sato-u-ss318_442_1999,sato-u-ss494_493_2001},
does not occur  in our previous simulations  for  impurity-induced step bunching~\cite{msato-jpsj2017,ms2018pre}.
We suggest that the separation of single steps may occur repeatedly
when we change the lifetime of impurities  $\tau_\mathrm{imp}$
or take into account  the  surface diffusion of impurities.
Therefore,  in our study,
we performed  Monte Carlo simulations  
and studied the  effects  of both  the  surface diffusion  of impurities
and the lifetime of impurities 
on the separation of single steps.
In Sec.~\ref{sec:model}, we introduce  our model,
in which we  add the  surface diffusion  of impurities to our previous model~\cite{ms2018pre}.
In Sec.~\ref{sec:results}, we show the results of our simulations.
In Ref.~\onlinecite{Vladimirova01prb245420}, the difference in  the Ehrich-Schwoebel barrier~\cite{Ehrlich-H,Schwoebel-s}
between dimers and monomers causes step bunching during growth.
In that study, the distribution of terrace width changes with 
the variation in the  impingement rate of monomers, 
the coefficient of
 diffusion of monomers, and the average terrace width on a  vicinal face.
Although the cause of  step bunching is different in our model,
we show that the distribution of terrace widths changes by controlling parameters even in our system.
In Sec.~\ref{sec:subsec1}, 
we show  the dependence of the form of step bunches on $\tau_\mathrm{imp}$.
In Sec.~\ref{sec:subsec2}, 
we show the relationship between the separation of steps and the frequency of 
the surface  diffusion of impurities.
In Sec.~\ref{sec:summary}, we summarize our  results.

\section{Model}\label{sec:model}
In general, kinetic Monte Carlo simulations are  based on implementing a set of events with given rates. 
We just have to  perform  the events  according to the rates.
However, we adopted  another approach used in previous studies~\cite{Uwaha-Saito92prl224,Uwaha-Saito93ss366,
Uwaha-Saito93jcg82,Saito-u94prb10677}.
This approach is not effective 
but it is easy to consider the correspondence with 
the Burton--Cabrera--Frank (BCF) model~\cite{Frank}.
We discretize the diffusion equation of adatoms and consider particles hopping on a lattice, which correspond
to adatoms.
Hopping of particles, and  solidification and melting  at steps occur  probabilistically.
Because solid atoms and adatoms are distinguished in the model, 
we can define the step stiffness $\tilde{\beta} $ and the equilibrium adatom density $c_\mathrm{eq}$ independently. 
We modify  the models adopted in our previous studies~\cite{msato-jpsj2017,ms2018pre} 
and conduct Monte Carlo simulations.

We consider  a square lattice, in which the lattice constant is unity.
The system lengths in the $x$ and $y$-directions are denoted 
 by $L_x$ and $L_y$, respectively.
We use a periodic boundary condition in the $x$-direction and a
 helical boundary condition
in the $y$-direction.
Initially,  straight $N$  steps are  set   parallel to the $x$-axis.
The steps advance to the $y$-direction during growth.
For simplicity,
we forbid two-dimensional nucleation on terraces.
\begin{figure}[htbp]
\centering

\includegraphics[width=8.0cm,clip]{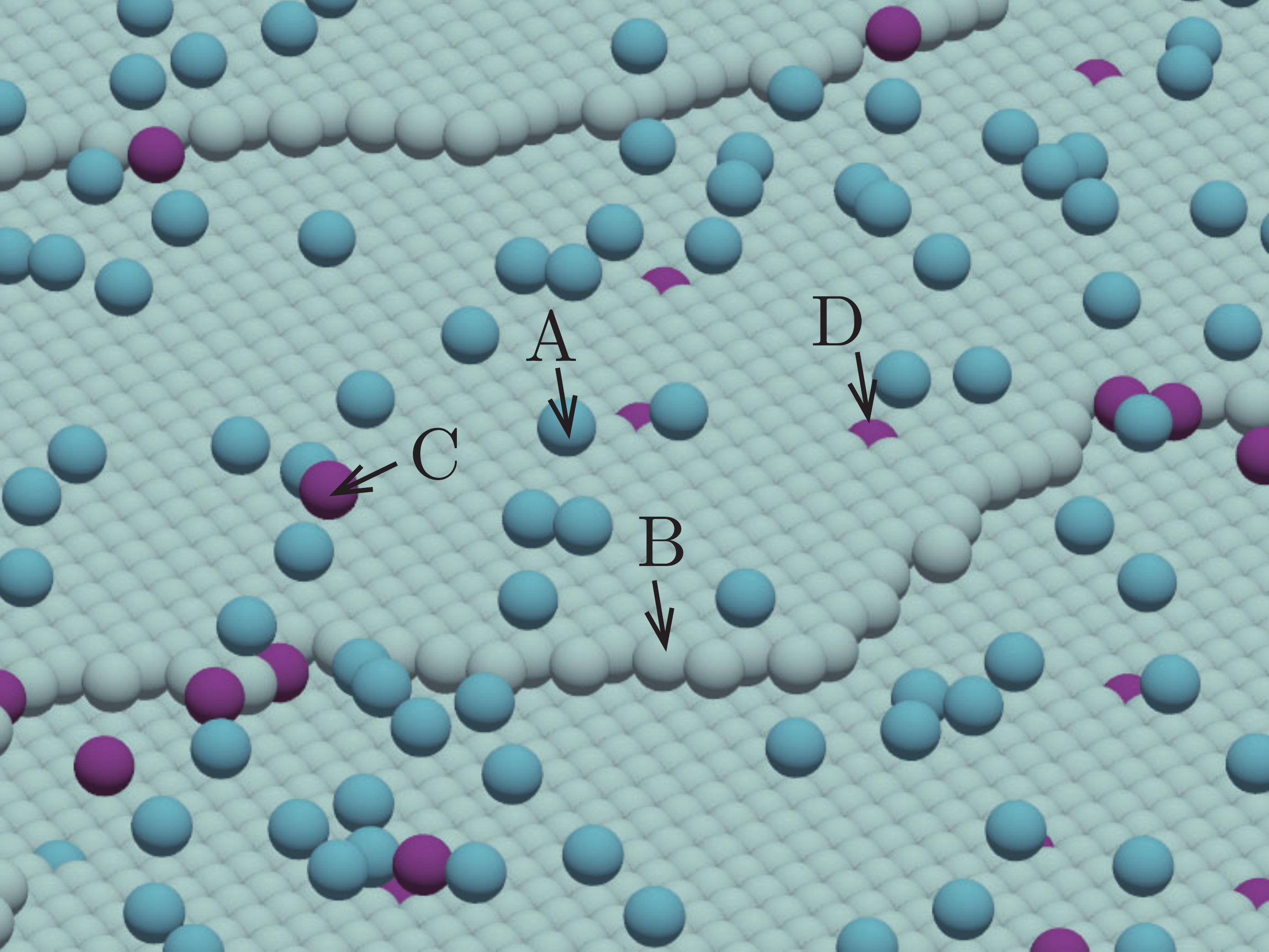} 

\caption{
(color online)
Schematic  of our model.
Particles (A), (B), (C), and (D) are an adatom,  a solid atom forming step,
an impurity on a surface, and an impurity incorporated into a  solid, respectively. 
}
\label{fig:schematic model}
\end{figure}

In our simulation, we choose adatoms, impurities on surface, or  solid atoms having  empty 
neighboring sites 
in the horizontal direction,
and try surface diffusion  of both adatoms and impurities, solidification of adatoms,
or melting of solid atoms.
When we  choose either an adatom   such as particle (A) in Fig.~\ref{fig:schematic model}
or an impurity  on the surface 
such as  particle (C) in Fig.~\ref{fig:schematic model},  
we first try  the  evaporation of the chosen particle.
If it  does not evaporate, we next try its  surface diffusion.
As  a  surface diffusion trial of an adatom,
we try to move  an adatom to one of  its four neighboring sites,
where the moving  probability is  $1/4$ for each site.
When  the selected site 
is already occupied by an impurity or an adatom, 
we do not move the selected adatom  and make it  stay at the same site. 
Because we set the coefficient of diffusion  for adatoms $D_\mathrm{s}$ to  unity,
the time increase in a  diffusion  trial of an adatom is 
given by $1/4N_\mathrm{s}$, where $N_\mathrm{s}$ 
is the number of adatoms.
We assume that  the coefficient of the surface diffusion of impurities 
$D_\mathrm{imp}$ is smaller than  $D_\mathrm{s}$.
The rule for the surface diffusion of impurities  is the same as that of adatoms
except that we determine whether we try the surface diffusion of impurities 
with a probability $D_\mathrm{imp}/(4D_\mathrm{s})$. 
Because  the steps are permeable in our model~\cite{Sato-us00prb8452},
both adatoms and impurities can  diffuse   on the surface over the steps.

We solidify adatoms when they attach to a step after a  diffusion  trial.
The solidification probability for  adatoms $p_\mathrm{s}$ is given by 
\begin{equation}
p_\mathrm{s}
= \left[ 1 + \exp\left( \frac{\Delta E - \phi}{k_\mathrm{B}T }\right) \right]^{-1},
\end{equation}
where $\Delta E$ represents the change in the step energy, $\phi$ the change in 
the  chemical potential  by solidification per an adatom,
$k_\mathrm{B}$  the Boltzmann constant,  and $T$ temperature.
When an adatom attaches  to $n_\mathrm{n}$  solid atoms 
 in the horizontal direction, $\Delta E$ is given by $2 \epsilon (2-n_\mathrm{n})$,
 where $\epsilon $ is the  bonding energy per one bond.
 For simplicity, we forbid the formation of multi-height steps.

 We do not solidify impurities
 but regard  that 
impurities as  incorporated into the solid  phase 
when  all their neighboring sites are occupied 
by  solid atoms and impurities  such as    particle (D) in Fig.~\ref{fig:schematic model}.
The impurities incorporated into the solid phase 
cannot  diffuse on a surface  nor evaporate  unless one or more  of the neighboring sites get empty.
We assume that
the surface diffusion of both adatoms and impurities on the  impurities incorporated
 into
the solid phase is possible.

The lifetimes of adatoms  and impurities on a vicinal face
are given  by $\tau$ and $\tau_\mathrm{imp}$,
respectively.
They are related to the evaporation probability of adatoms $p_\mathrm{eva} $ and that of 
impurities $p_\mathrm{eva}^\mathrm{imp}$ as $p_\mathrm{eva}  = 1/4\tau$ and 
$p_\mathrm{eva}^\mathrm{imp}  = 1/4\tau_\mathrm{imp}$.
After some evaporation and diffusion trials,
impurities and  atoms  impinge on the sites, which are not 
occupied by other particles, at random
with the impingement rates  $F_\mathrm{imp}$ and $F$, respectively.
When impurities are contained in raw materials,
the ratio of $F_\mathrm{imp}$ to $F$ is probably kept constant  even if $F$changes. 
Therefore, we  performed  simulations keeping  $F_\mathrm{imp}/F$ constant


We  also try melting
of solid atoms, for example  particle (B) in Fig.~\ref{fig:schematic model},
which is a solid atom having an empty neighboring site  at least  and no adatom on it.
When a melting trial succeeds,  the melted atom stays on the same site
as an adatom.
The melting probability $p_\mathrm{m}$ is given by 
\begin{equation}
p_\mathrm{m}
= \left[ 1 + \exp\left( \frac{\Delta E +\phi}{k_\mathrm{B}T }\right) \right]^{-1}.
\end{equation}
The frequencies of solidification and melting trials are the same at kink sites, where $\Delta E=0$
in equilibrium.  Because the equilibrium adatom density $c_\mathrm{eq}$ satisfies 
$c_\mathrm{eq} p_\mathrm{s} =(1-c_\mathrm{eq})p_\mathrm{m}$ at the kink sites,
$c_\mathrm{eq}$ is given by~\cite{Saito-u94prb10677}
\begin{equation}
c_\mathrm{eq} = 
\left[ 
1 + \exp \left( \frac{\phi}{k_\mathrm{B}T } \right ) 
\right]^{-1} .
\end{equation}
For  small step fluctuations,
the step stiffness $\tilde{\beta}$, 
 which  represents the increase in the step free energy by step fluctuation,
can be estimated in our model.
When we assume that the step position $y$ is a single function of $x$ because of a small step fluctuation,
the difference in the step position between neighboring sites $n=(y(x_{i+1})-y(x_{i}))/a$
is related to the step stiffness as~\cite{Saito-u94prb10677} 
\begin{equation}
\tilde{\beta} =\frac{k_\mathrm{B}T}{a \langle n^2 \rangle}=
\frac{2 k_\mathrm{B}T}{a}	
\sinh^2 \frac{\epsilon}{2k_\mathrm{B}T},
\label{eq:stiffness}
\end{equation}
where $\langle n^2 \rangle$ is the ensemble average of  $n^2$.
Because $\langle n^2 \rangle$ is as large as the average kink density for small step fluctuations,
small $\tilde{\beta} $ means the formation of many kinks.

\section{Results of simulations}\label{sec:results}
We set the system size $L_x \times L_y$ to  $ 512\times 512$
and the number of steps $N$ to 16 in our simulations.
Initially,  the steps are straight and equidistant.
The initial number of adatoms is  roughly equal to that in equilibrium and impurities are not present on the vicinal face.
We set
$\epsilon/k_\mathrm{B}T=2.0$ and $\phi/k_\mathrm{B}T=3.0$,
which are the same as those used in our previous study~\cite{ms2018pre}.
Because  the capillary length, which is given by $a^2\tilde{\beta}/(k_\mathrm{B}T)$, 
is estimated to be $2.76$, there are many kinks on the steps.
The adatom density under an equilibrium condition is low in the simulation system because $c_\mathrm{eq}$
is set to $4.7 \times 10^{-2}$.
Other parameters,
$F$, $F_\mathrm{imp}/F$, and $\tau$ 
are set to  $ 1.4 \times 10^{-3}$ , $ 4 \times 10^{-3}$,
and $ 512$,  respectively.



\subsection{Dependence of step bunching on the lifetime of impurities }\label{sec:subsec1}
In our previous study~\cite{ms2018pre}  with large lifetimes of impurities, 
stable step bunches are formed and the separation of single steps from bunches,
which occurs in other 
systems~\cite{sato-u-prb11172_51_1995,sato-u-ss318_442_1999,sato-u-ss494_493_2001},
is not observed.
We think that the separation of steps
may repeatedly occur when $\tau_\mathrm{imp}$ is small.
Hence,
we neglect the surface diffusion of impurities for simplicity
and study  how  behaviors of step bunches change  by  decreasing  $\tau_\mathrm{imp}$.

\begin{figure}[htbp]
\centering

\includegraphics[width=5.5cm,clip]{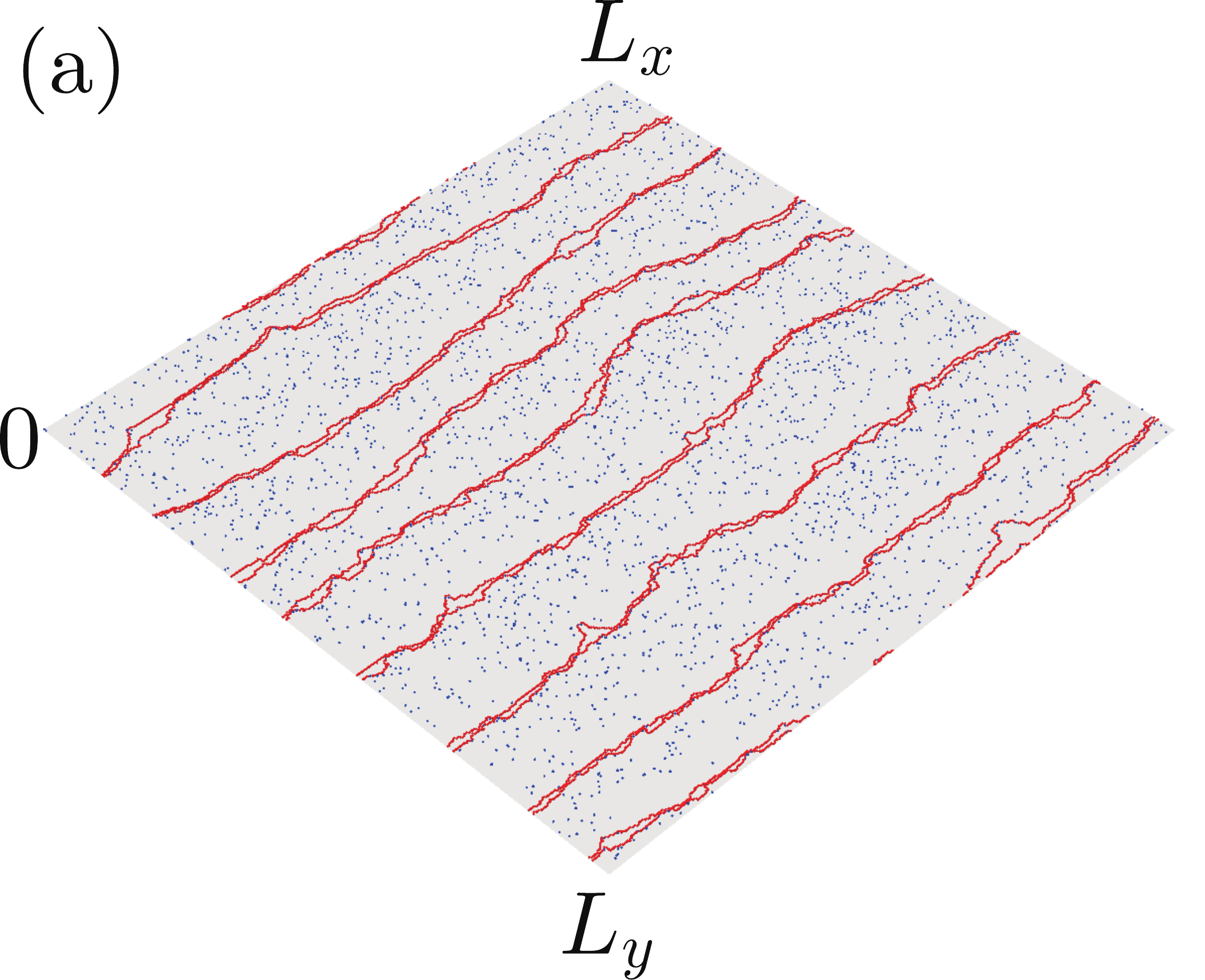}
\includegraphics[width=5.5cm,clip]{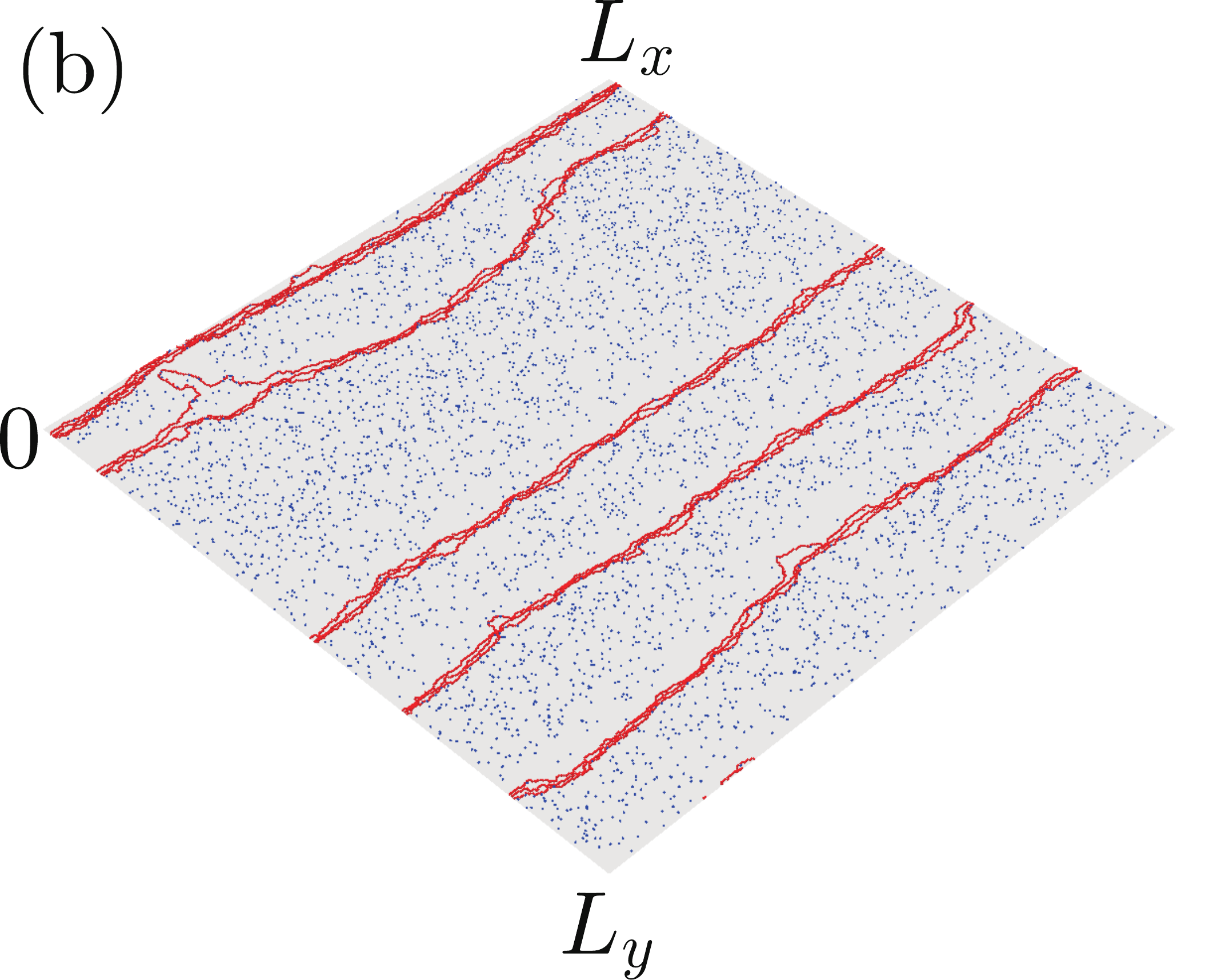} 
\vspace*{0.5cm}

\includegraphics[width=5.5cm,clip]{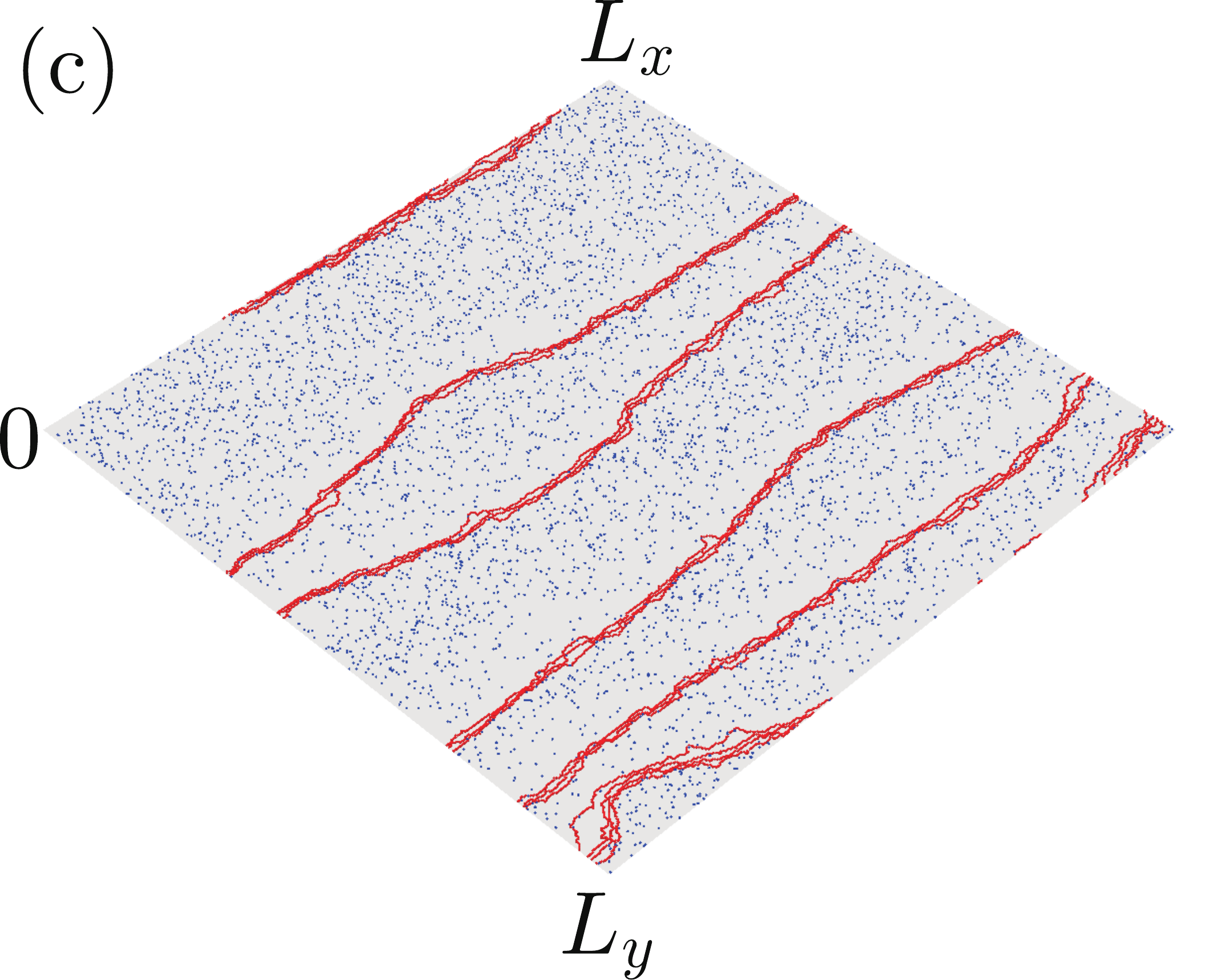} 
\includegraphics[width=5.5cm,clip]{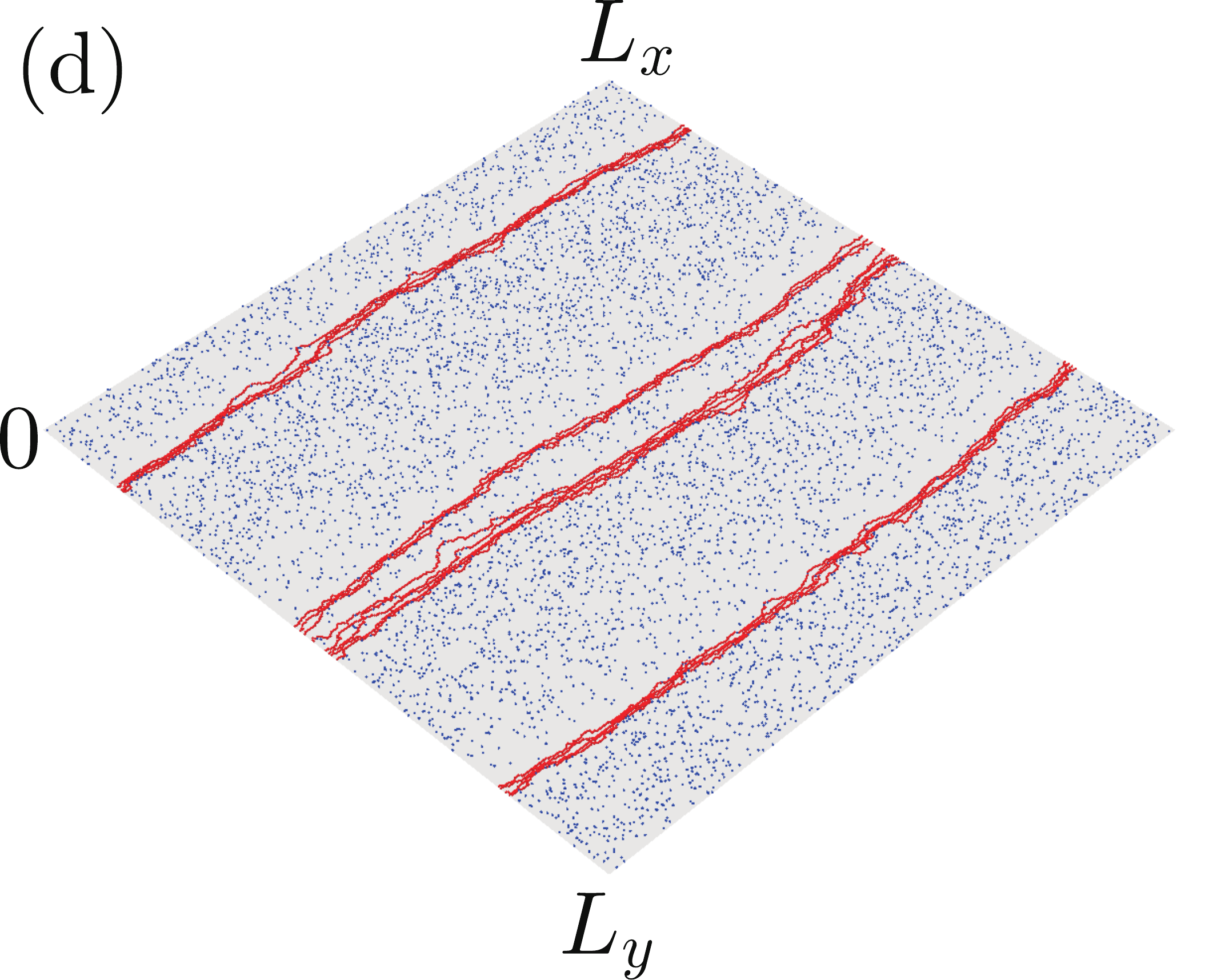} 
\vspace*{0.5cm}

\includegraphics[width=5.5cm,clip]{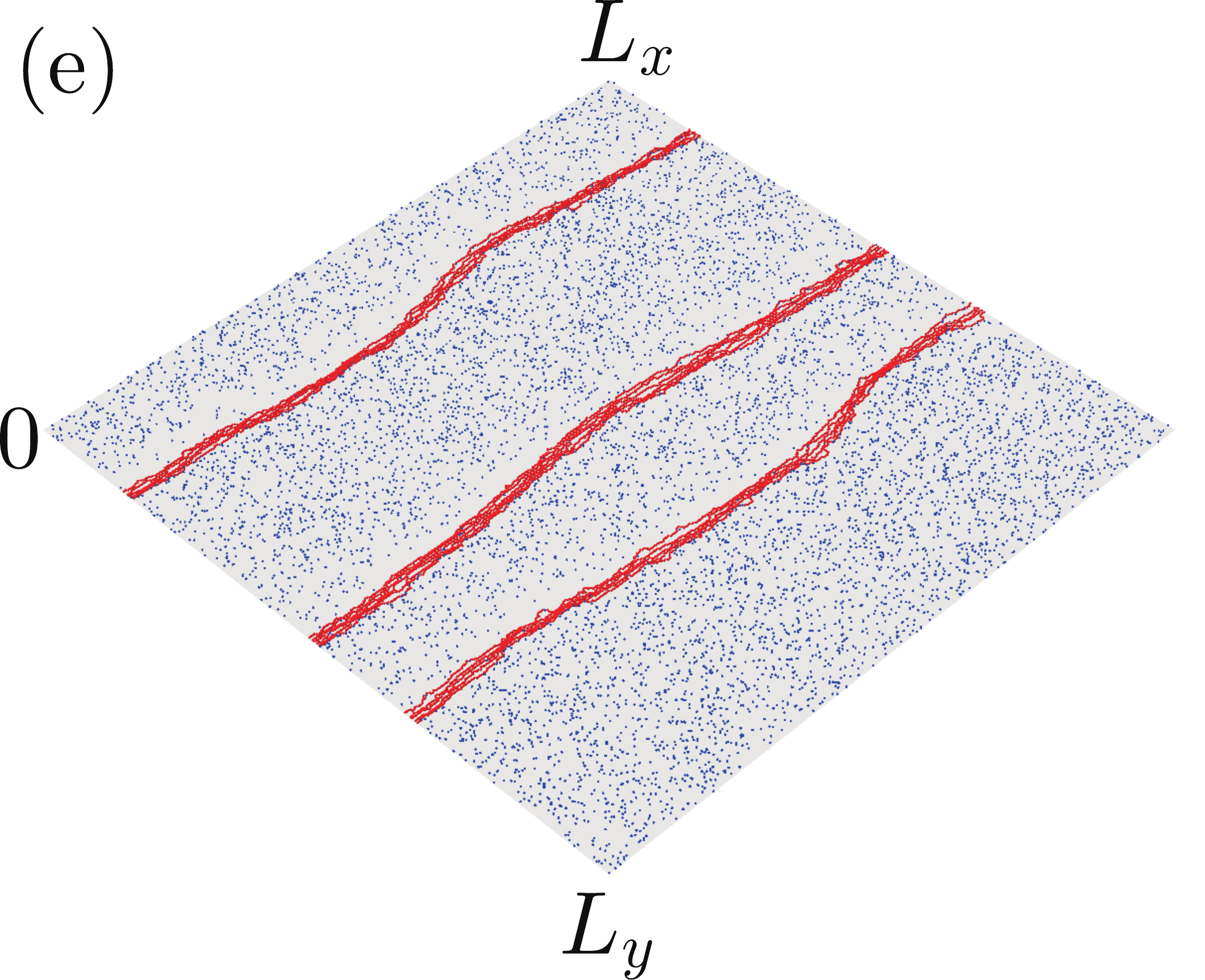} 
\includegraphics[width=5.5cm,clip]{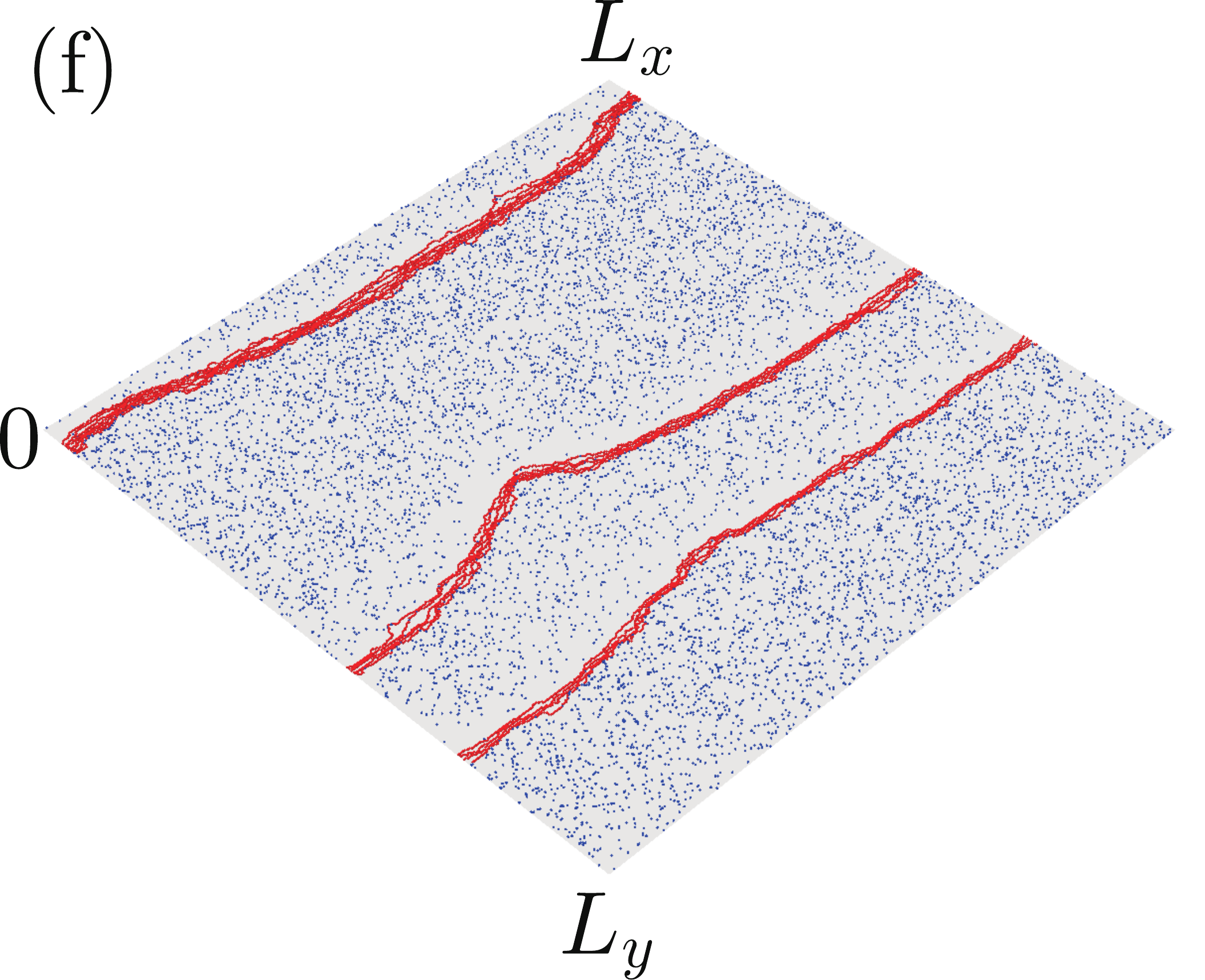}

\caption{
(color online)
Snapshots  of step bunches with 
(a) 
$\tau_\mathrm{imp}= 3 \times 10^4$ at  $t = 3.7 \times 10^6$,
(b) 
$\tau_\mathrm{imp}= 4 \times 10^4$ at  $t = 3.5 \times 10^6$,
(c) 
$\tau_\mathrm{imp}= 5 \times 10^4$ at  $t = 3.4 \times 10^6$,
(d) 
$\tau_\mathrm{imp}= 6 \times 10^4$ at  $t = 3.3 \times 10^6$,
(e) 
$\tau_\mathrm{imp}= 7 \times 10^4$ at  $t = 3.1 \times 10^6$,
and 
(f) 
$\tau_\mathrm{imp}= 8 \times 10^4$ at  $t = 3.1 \times 10^6$,
where we set $D_\mathrm{imp}/D_\mathrm{s}$ to $0$.
The blue dots on the surface represent impurities.}
\label{fig:snapshot_tau80000}
\end{figure}
Figure~\ref{fig:snapshot_tau80000} shows  snapshots  of  surfaces in  late stages of 
step bunching
for various $\tau_\mathrm{imp}$,
where  large bunches are formed and no single steps are seen on large terraces.
The bunch size $N_\mathrm{B}$, which means the number of steps in a bunch, decreases with decreasing $\tau_\mathrm{imp}$.
As  step bunches are  straight 
in each case,  we average the step positions in  the $x$-direction
and see the time evolution of the average step positions.

\begin{figure}[htbp]
\centering

\includegraphics[width=6.0cm,clip]{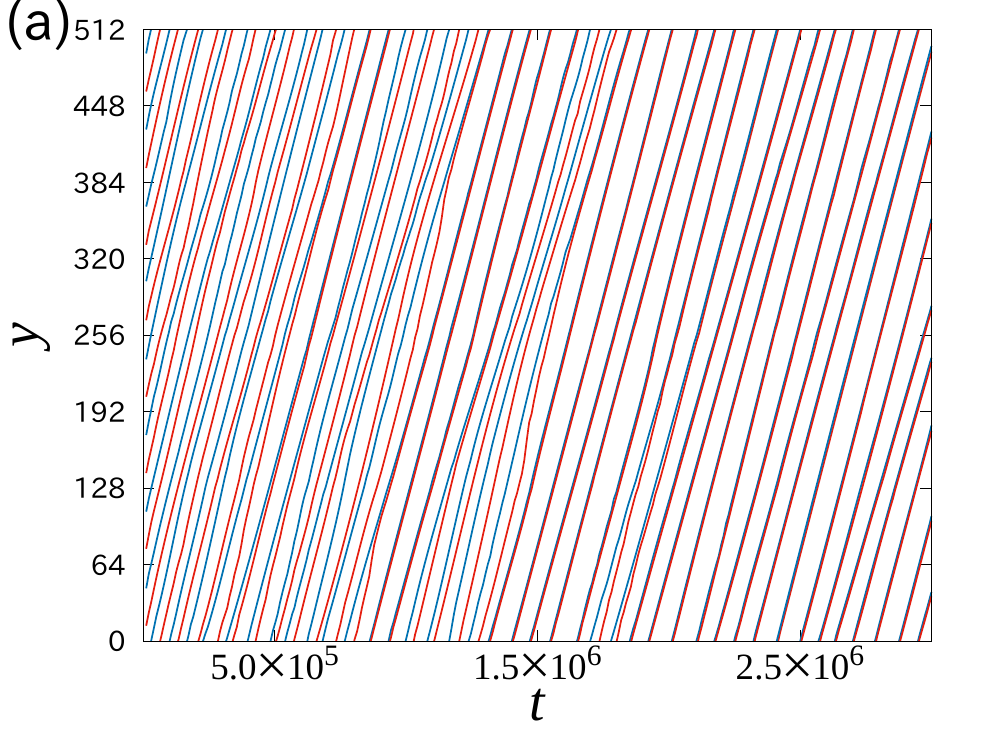} 
\includegraphics[width=6.0cm,clip]{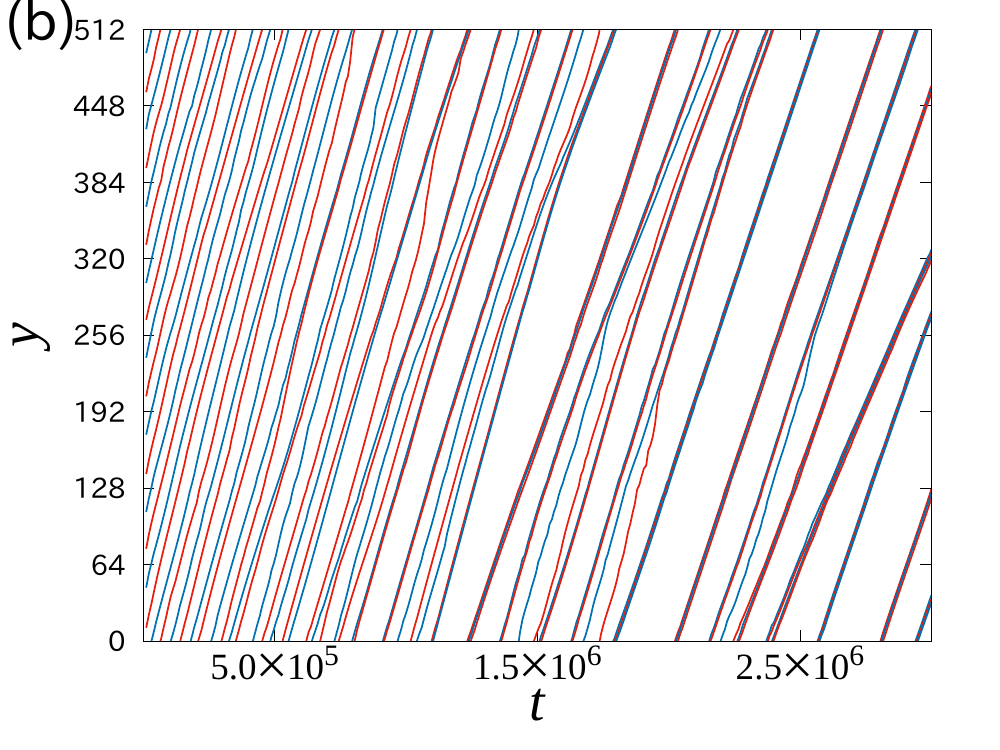} 

\includegraphics[width=6.0cm,clip]{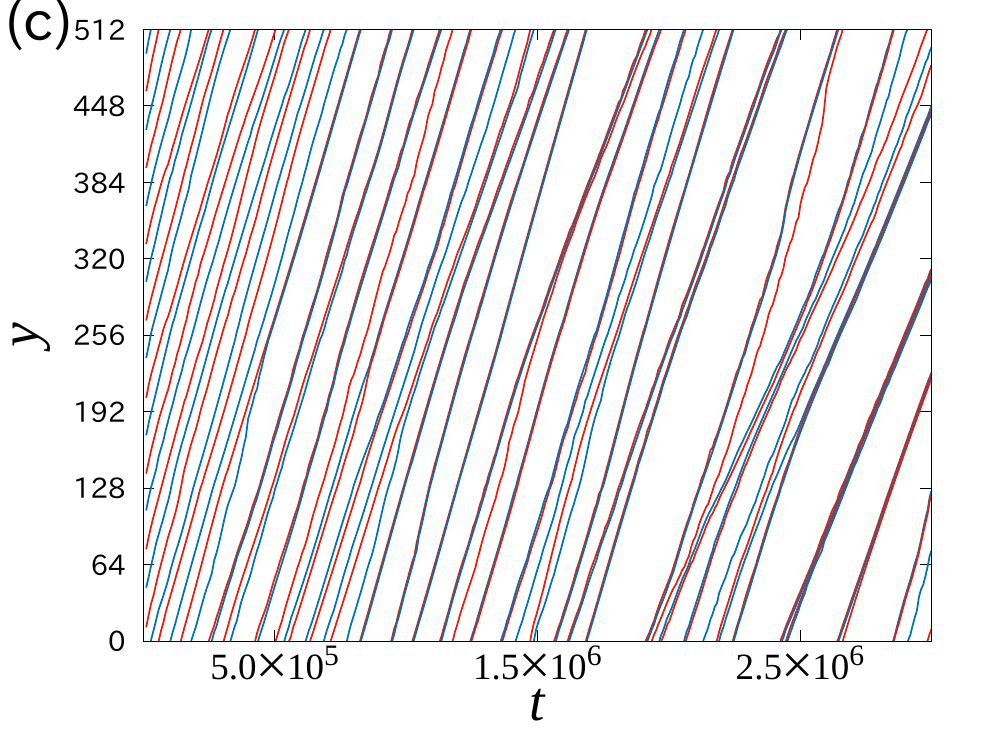} 
\includegraphics[width=6.0cm,clip]{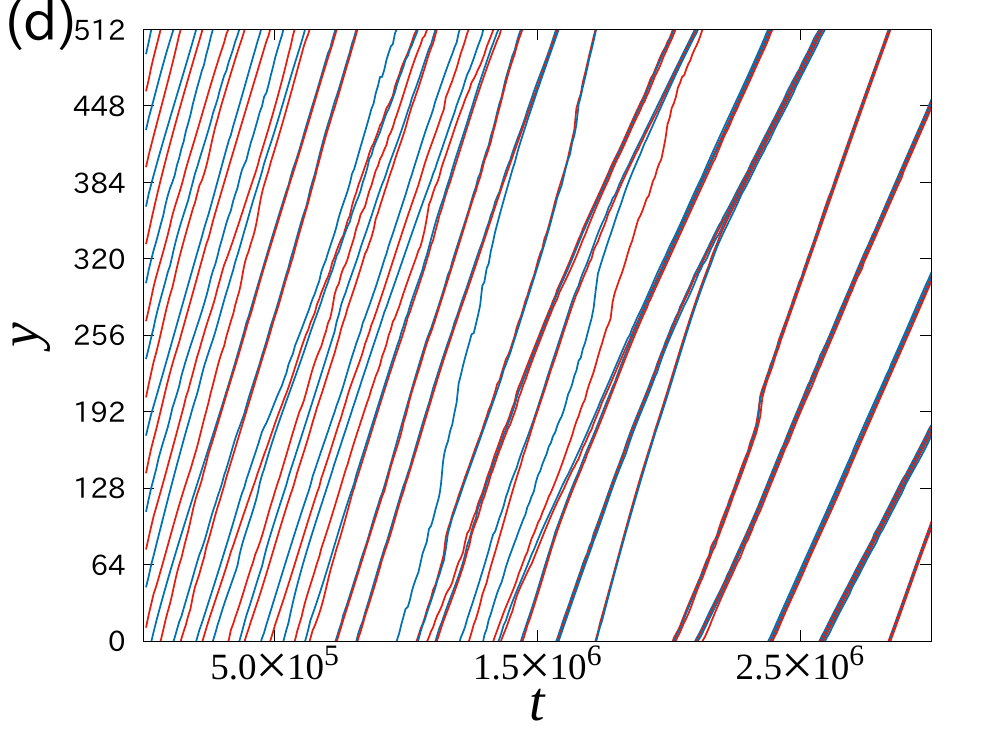} 

\includegraphics[width=6.0cm,clip]{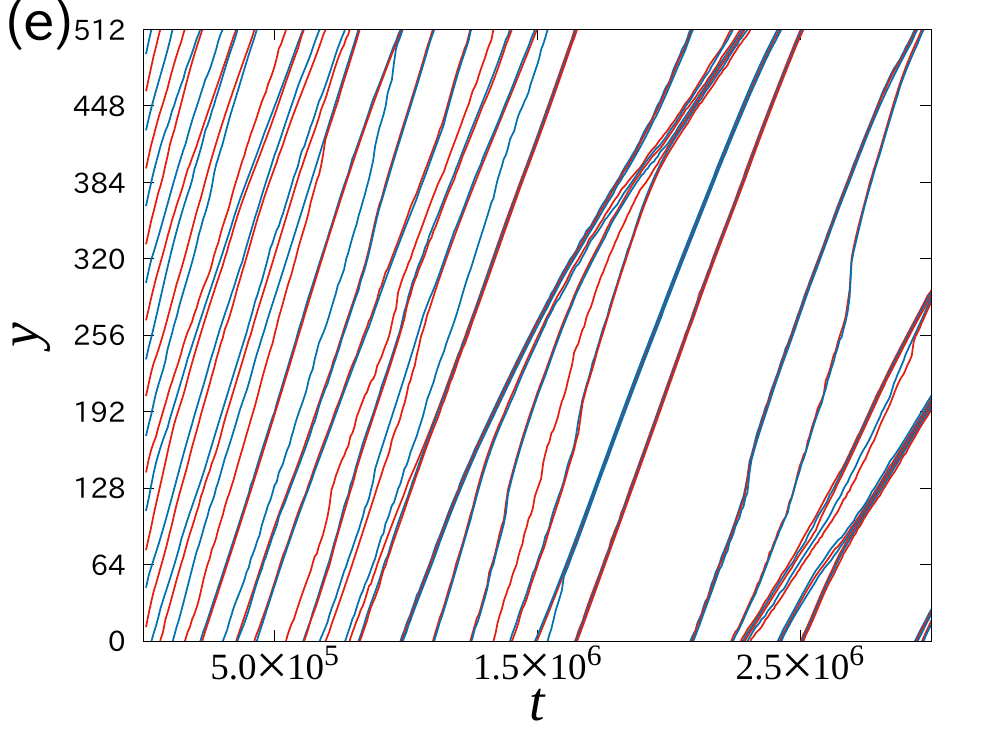} 
\includegraphics[width=6.0cm,clip]{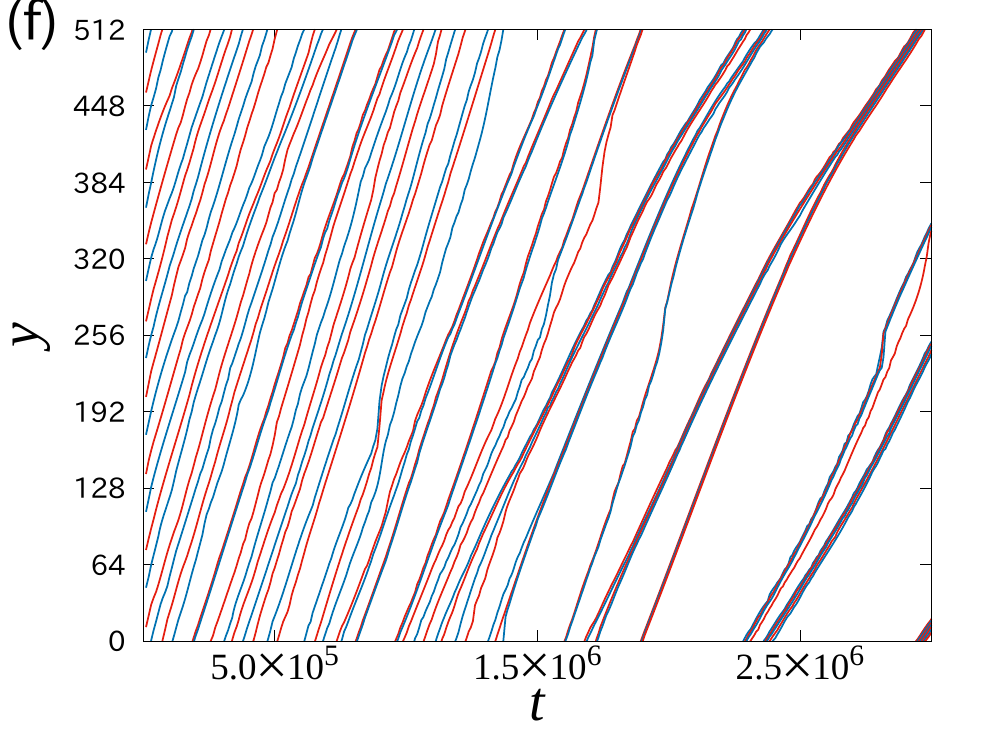}

\caption{
(color online)
Time evolution of step positions averaged in the $x$-direction:
(a)--(f) 
for the sample used in Figs.~\ref{fig:snapshot_tau80000}(a)--(f),
respectively.
$\tau_\mathrm{imp}$ 
is 
(a) $3 \times 10^4$,
(b) $4 \times 10^4$,
(c) $5 \times 10^4$,
(d) $6 \times 10^4$,
(e) $7 \times 10^4$,
and 
(f) $8  \times 10^4$.}
\label{fig:timeevolution_tau80000}
\end{figure}
Figures~\ref{fig:timeevolution_tau80000}(a)--(f)  show the time evolution  of 
average step positions  for the samples used in Figs.~\ref{fig:snapshot_tau80000}(a)--(f), respectively.
The initial equispaced steps are  unstable against step fluctuations and small 
bunches  form
in the  early stages.
The bunch size $N_\mathrm{B}$ increases  with  the collision of small bunches.
Single steps temporarily appear  because of   the collisions of bunches in 
the early stages,
but the single steps are rarely seen in  the late stage.
Because we might see rare cases in Figs.~\ref{fig:snapshot_tau80000} and \ref{fig:timeevolution_tau80000},
we try  ten individual runs for each $\tau_\mathrm{imp}$
and study how the number of isolated steps $N_\mathrm{iso}$  decreases and $N_\mathrm{B}$ increases with  time.

\begin{figure}[htbp]
\centering

\includegraphics[width=8.0cm,clip]{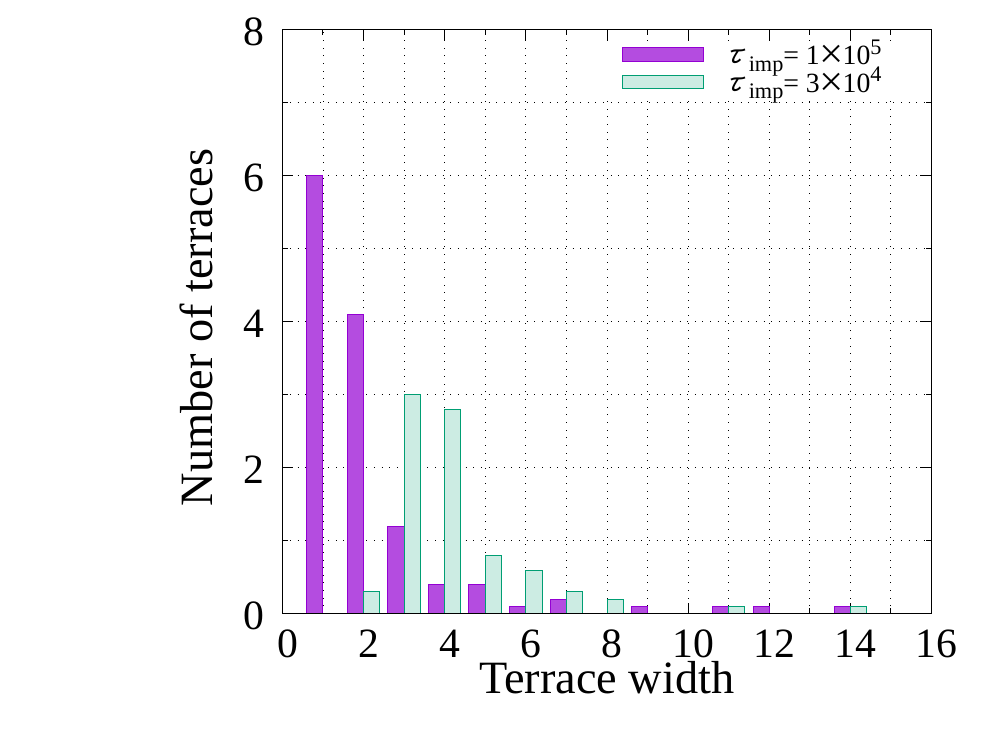} 

\caption{
(color online)
Distributions of terrace widths  that  are smaller than 16
for $\tau_\mathrm{imp}= 3 \times 10^4$ and 
$\tau_\mathrm{imp}= 1 \times 10^5$.
The data are averaged over 10 individual runs.}
\label{fig:terrace_width}
\end{figure}
Figure~\ref{fig:terrace_width} shows the distributions 
of terrace widths  that  are smaller than  16  for $\tau_\mathrm{imp}= 3 \times 10^4$ and 
$1 \times 10^5$.
The number of small terraces increases with increasing $\tau_\mathrm{imp}$
As $N_\mathrm{B}$   seems to  increase with increasing $\tau_\mathrm{imp}$
in Fig~\ref{fig:snapshot_tau80000}, 
the bunch becomes tight with increasing $N_\mathrm{B}$.
We regard steps as isolated when their upper side terrace  and lower side terrace are longer than a critical width $l_\mathrm{c}$.
Because the number of terraces  for which the  width is wider than six is small in Fig.~\ref{fig:terrace_width},
we set $l_c$ to six hereinafter.
\begin{figure}[htbp]
\centering

\includegraphics[width=8.0cm,clip]{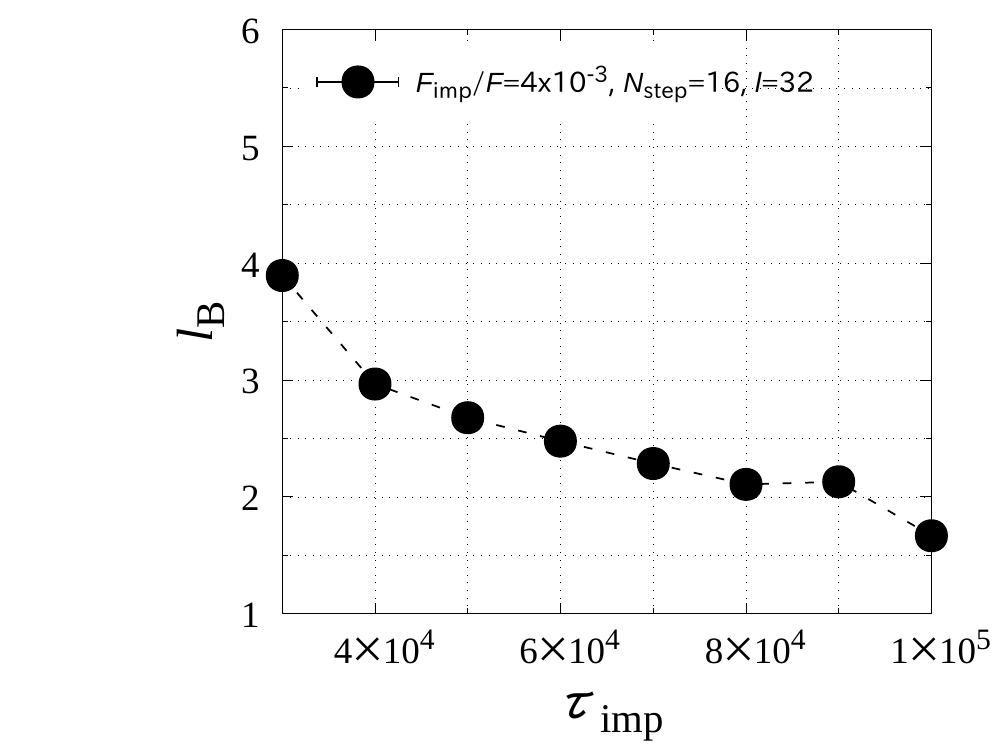} 

\caption{
Dependence of $l_\mathrm{B}$   on $\tau_\mathrm{imp}$.
The data are averaged over 10 individual runs.}
\label{fig:ave-l}
\end{figure}
When we use this criterion,
the average step distance in bunches $l_\mathrm{B}$ decreases with increasing $\tau_\mathrm{imp}$  (Fig.~\ref{fig:ave-l}).

\begin{figure}[htbp]
\centering

\includegraphics[width=8.0cm,clip]{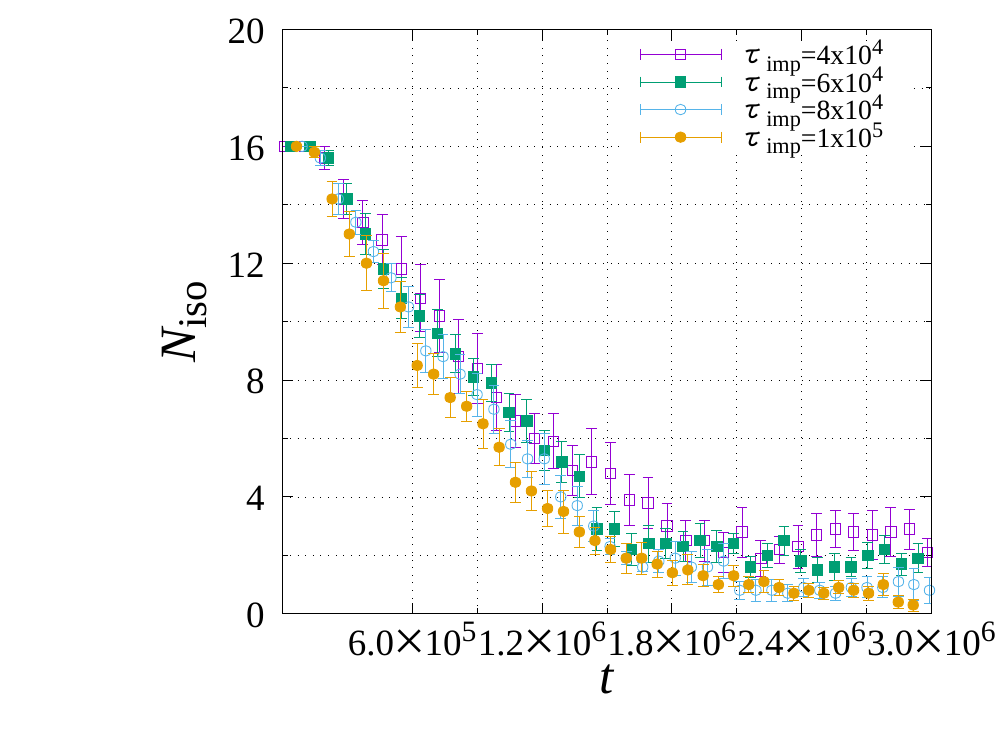} 

\caption{
(color online)
Time dependence of $N_\mathrm{B}$.
The data are averaged over 10 individual runs.}
\label{fig:number_singlesteps}
\end{figure}
Figure~\ref{fig:number_singlesteps} shows the time dependence of $N_\mathrm{iso}$.
With increasing time, 
$N_\mathrm{iso}$ decreases and  is just one or two in the systems in the last stage.
\begin{figure}[htbp]
\centering

\includegraphics[width=8.0cm,clip]{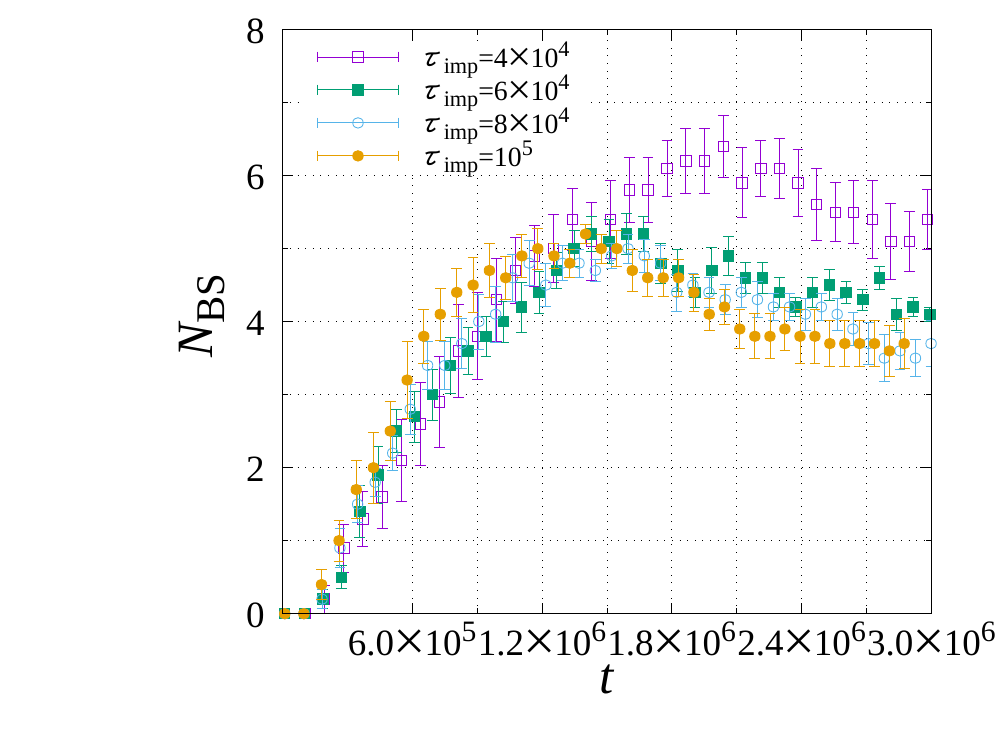} 

\caption{
(color online)
Time dependence of  $N_\mathrm{BS}$.
The data are averaged over 10 individual runs.}
\label{fig:bunch_size}
\end{figure}
Figure~\ref{fig:bunch_size} shows how the average number of bunches in systems
$N_\mathrm{BS}$ depends on time.
As single steps gather and small bunches forms,
$N_\mathrm{BS}$ increases with time in the initial stages.
However, $N_\mathrm{BS}$ decreases with time in the 
 later stages  because $N_\mathrm{B}$  increases 
from the collision of bunches
and $N_\mathrm{BS}$ seems to saturate in the last stage.

\begin{figure}[htbp]
\centering

\includegraphics[width=8.0cm,clip]{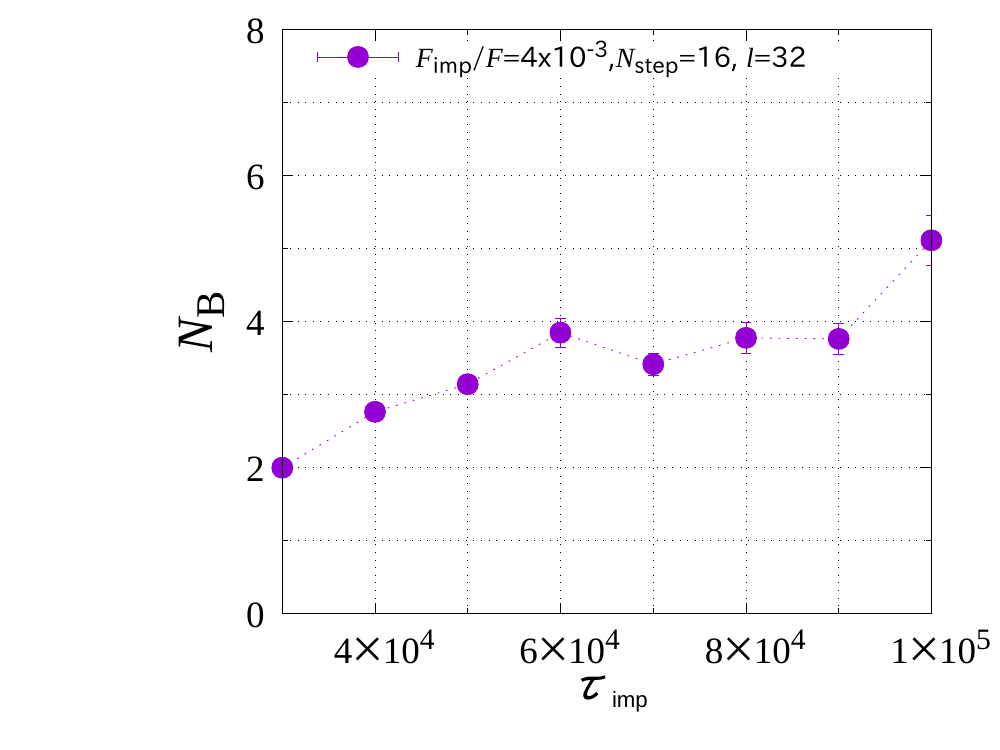} 

\caption{
(color online)
Dependence of $N_\mathrm{B}$  in the last stage.
The data are averaged over 10 individual runs.}
\label{fig:final_bunch_size}
\end{figure}
In Fig.~\ref{fig:final_bunch_size}, we show the dependence of  $N_\mathrm{B}$ in the last stage
on $\tau_\mathrm{imp}$.
Although the change in $N_\mathrm{B}$  
is small  in our simulation range of $\tau_\mathrm{imp}$ because of the limited system size,
$N_\mathrm{B}$  increases with increasing $\tau_\mathrm{imp}$.

%
\begin{figure}[htbp]
\centering

\includegraphics[width=8.0cm,clip]{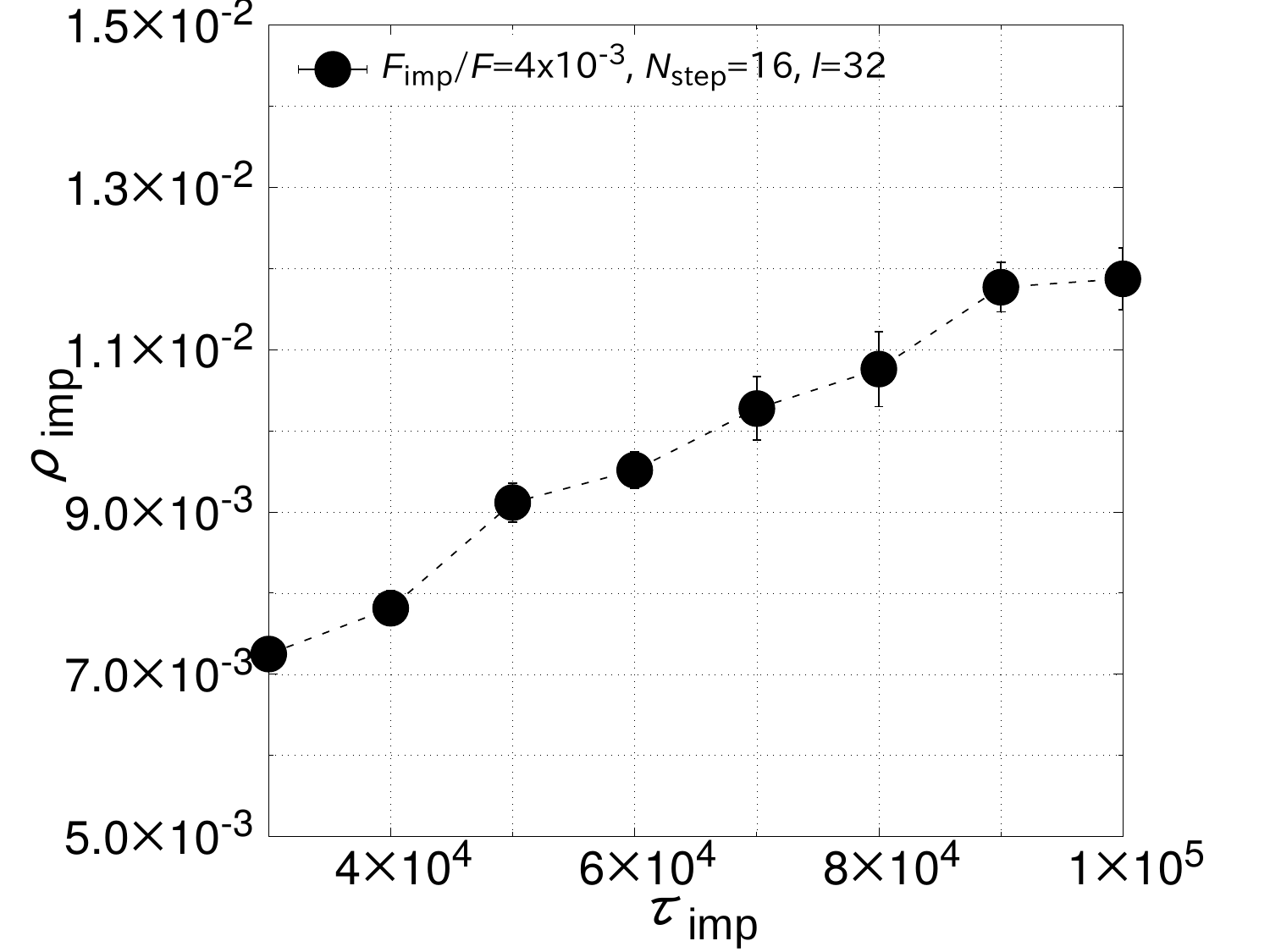} 

\caption{
Dependence of  $\rho_\mathrm{imp}$ 
on $\tau_\mathrm{imp}$ in the last stage.
The data are averaged over 10 individual runs.
}
\label{fig:impurity_insol}
\end{figure}
\begin{figure}[htbp]
\centering

\includegraphics[width=8.0cm,clip]{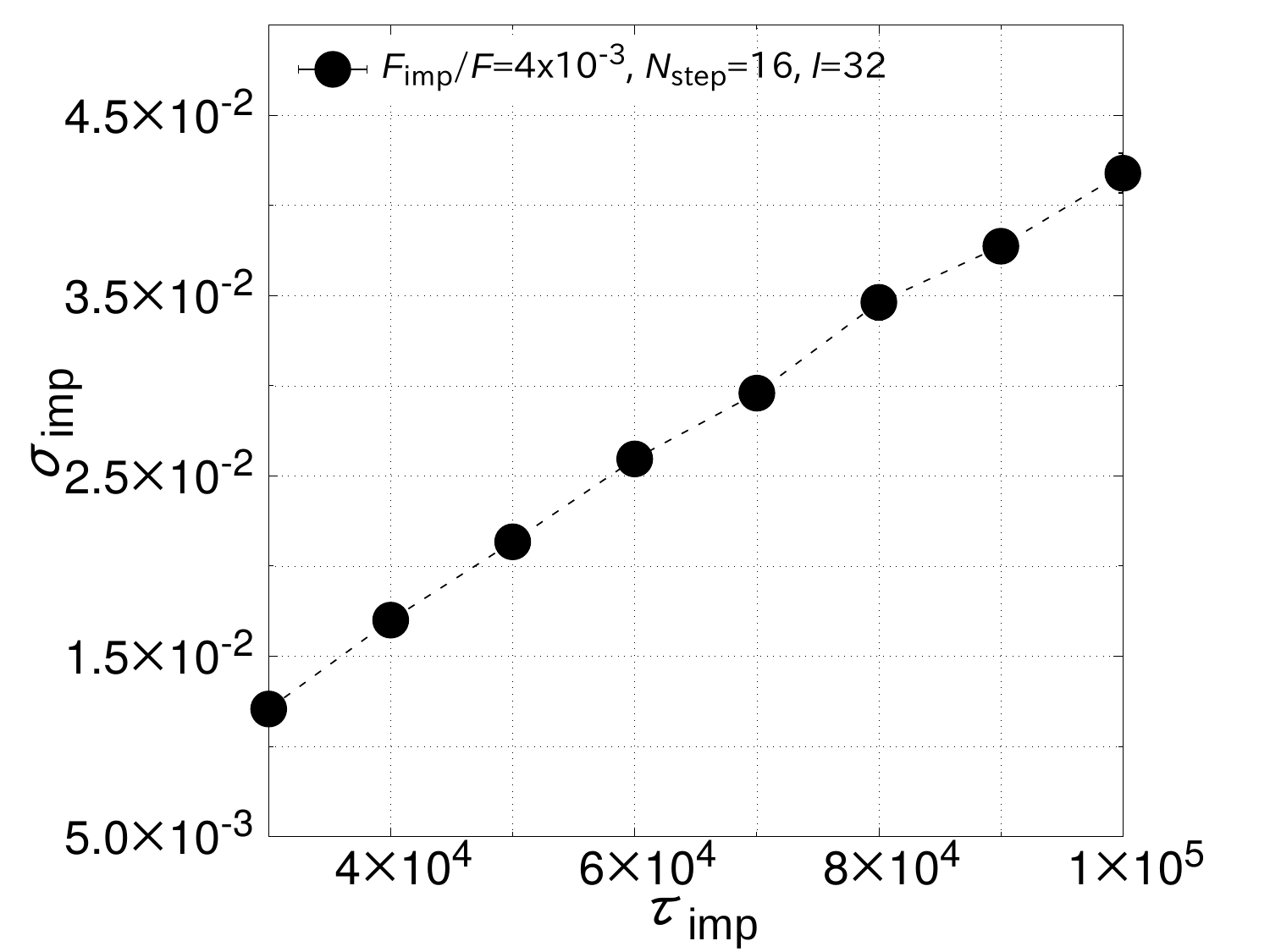} 

\caption{
Dependence of  $\sigma_\mathrm{imp}$
on $\tau_\mathrm{imp}$  in the last stage.
The data are averaged over 10 individual runs.
}
\label{fig:impurity_onsurface}
\end{figure}
The adatom density on surface $\sigma_\mathrm{imp}$ 
and the ratio of impurities incorporated into solid to solidified atoms  $\rho_\mathrm{imp}$
in the last stage depend on $\tau_\mathrm{imp}$;
see  Figs.~\ref{fig:impurity_insol} and~\ref{fig:impurity_onsurface}.
Both $\sigma_\mathrm{imp}$ and   $\rho_\mathrm{imp}$ increase with increasing $\tau_\mathrm{imp}$.
In our previous paper~\cite{ms2018pre},
we supposed  the impurity density $c_\mathrm{imp}$ obeys 
 \begin{equation}
 \frac{dc_\mathrm{imp} }{dt} = F_\mathrm{imp} -\frac{c_\mathrm{imp}}{\tau_\mathrm{imp}}.
 \end{equation}
We assumed that bunches with size $N_\mathrm{B}$ are formed equidistantly
and   that the separation of steps from bunches does not occur.
The incorporation of impurities in the solid phase occurs by advancing the lowest steps in 
 bunches.
 When the distance between bunches is longer than the surface diffusion length,
 $\rho_\mathrm{imp}$ is given by 
 \begin{equation}
 \rho_\mathrm{imp} =\frac{\Omega \tau_\mathrm{imp} F_\mathrm{imp}}{N_\mathrm{B}}
\left [
1-\exp \left( -\frac{N_\mathrm{B}^2 l }{\Omega x_\mathrm{s} (F-F_\mathrm{eq}) \tau_\mathrm{imp}} \right)
\right],
\label{eq:rho}
 \end{equation}
 where $\Omega$ is the atomic area,
 $l$ the step distance in a vicinal face,
 $x_\mathrm{s}$ the surface diffusion length defined as $x_\mathrm{s}= \sqrt{D_\mathrm{s}} \tau$,
$\tau$ the lifetime of adatoms,  
and $F_\mathrm{eq}$ is
given by $ c_\mathrm{eq}/\tau$.
In our simulations, 
$\tau=512$ and $D_\mathrm{s} = 1$,
so that $x_\mathrm{s}$  is estimated to be  $22.6$.
The prefactor in front of the parenthesis mainly changes $\rho_\mathrm{imp}$
because the exponential term on the right-hand side of Eq.~(\ref{eq:rho})
is negligibly small.
As
the change in $N_\mathrm{B}$ is small in our simulation,
the change in $\rho_\mathrm{imp}$  mainly is caused by  $\tau_\mathrm{imp}$.

\subsection{Dependence of step bunching on the  coefficients
of surface diffusion}\label{sec:subsec2}
Next,  we changed $D_\mathrm{imp}/D$
and  studied  how   surface diffusion of impurities
affects the step bunching induced by impurities.
\begin{figure}[htbp]
\centering

\includegraphics[width=5.5cm,clip]{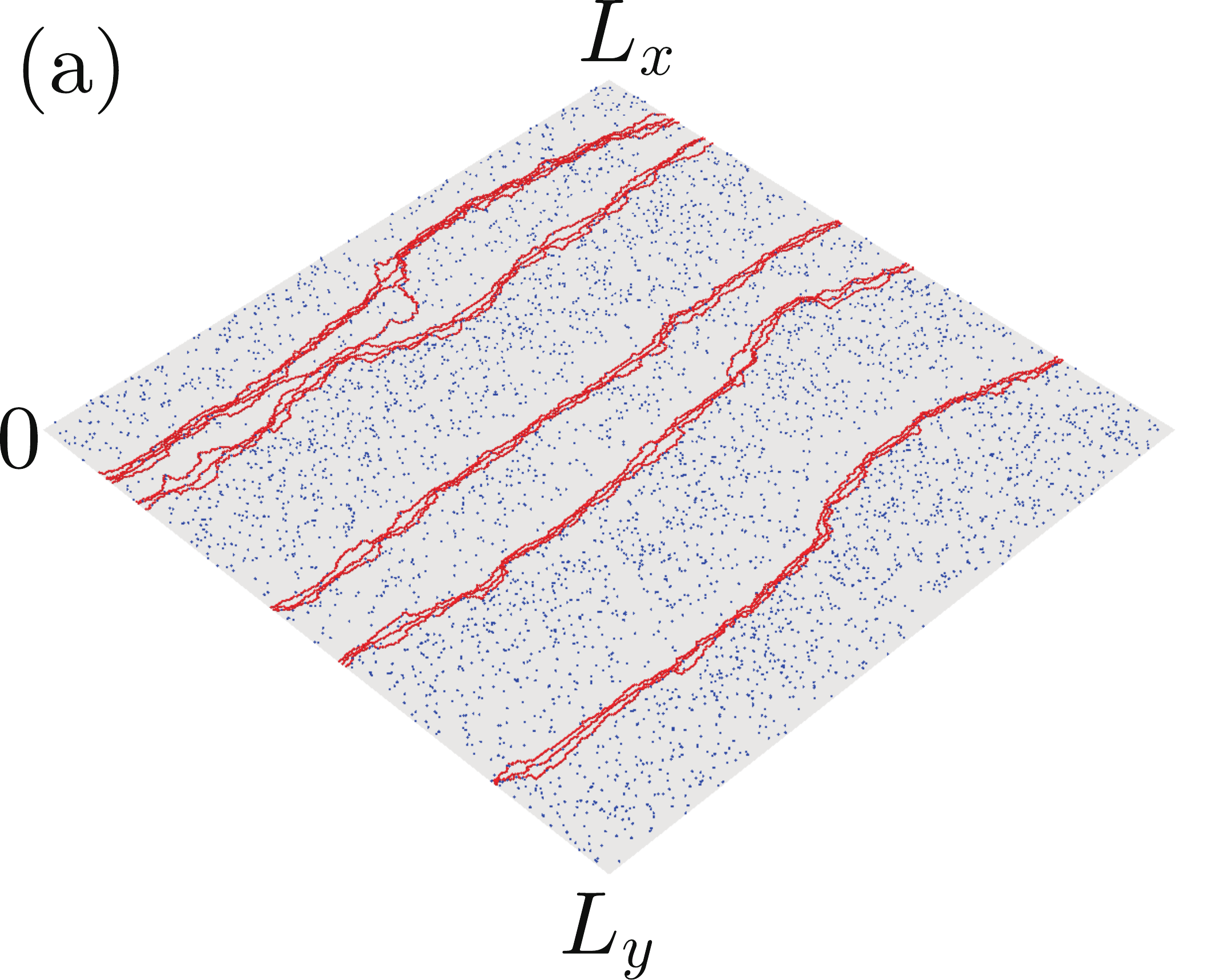} 
\includegraphics[width=5.5cm,clip]{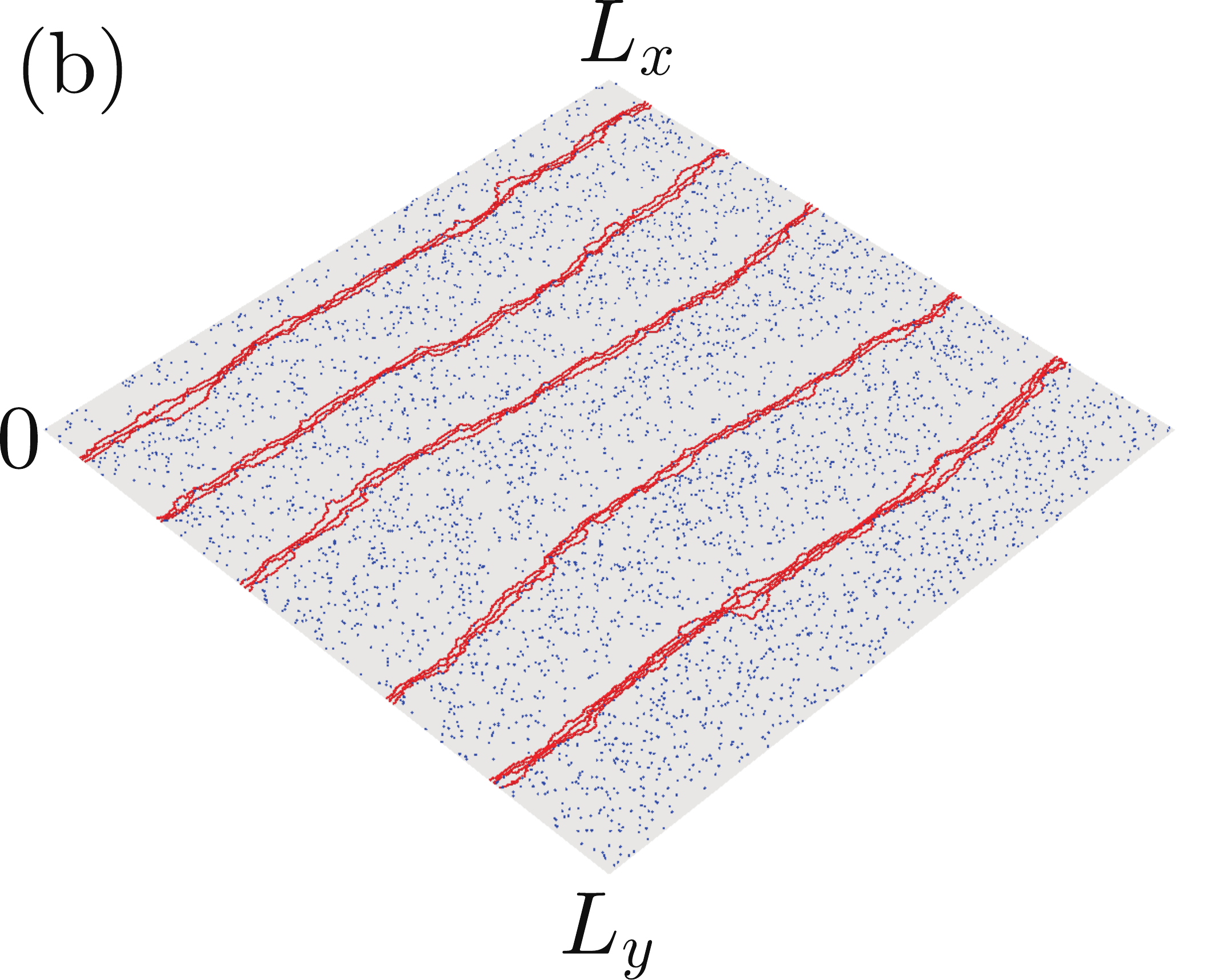} 
\vspace*{0.5cm}

\includegraphics[width=5.5cm,clip]{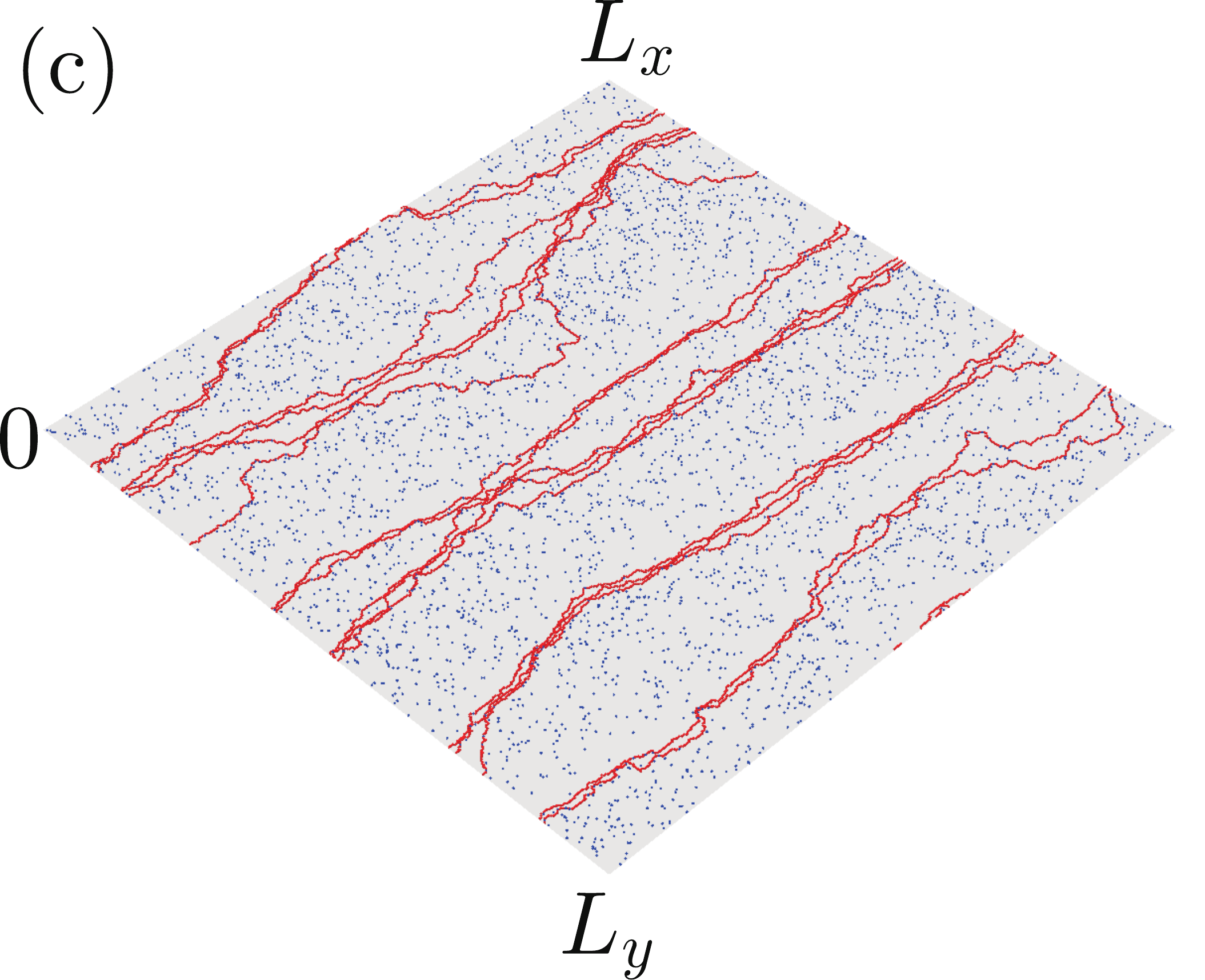} 
\includegraphics[width=5.5cm,clip]{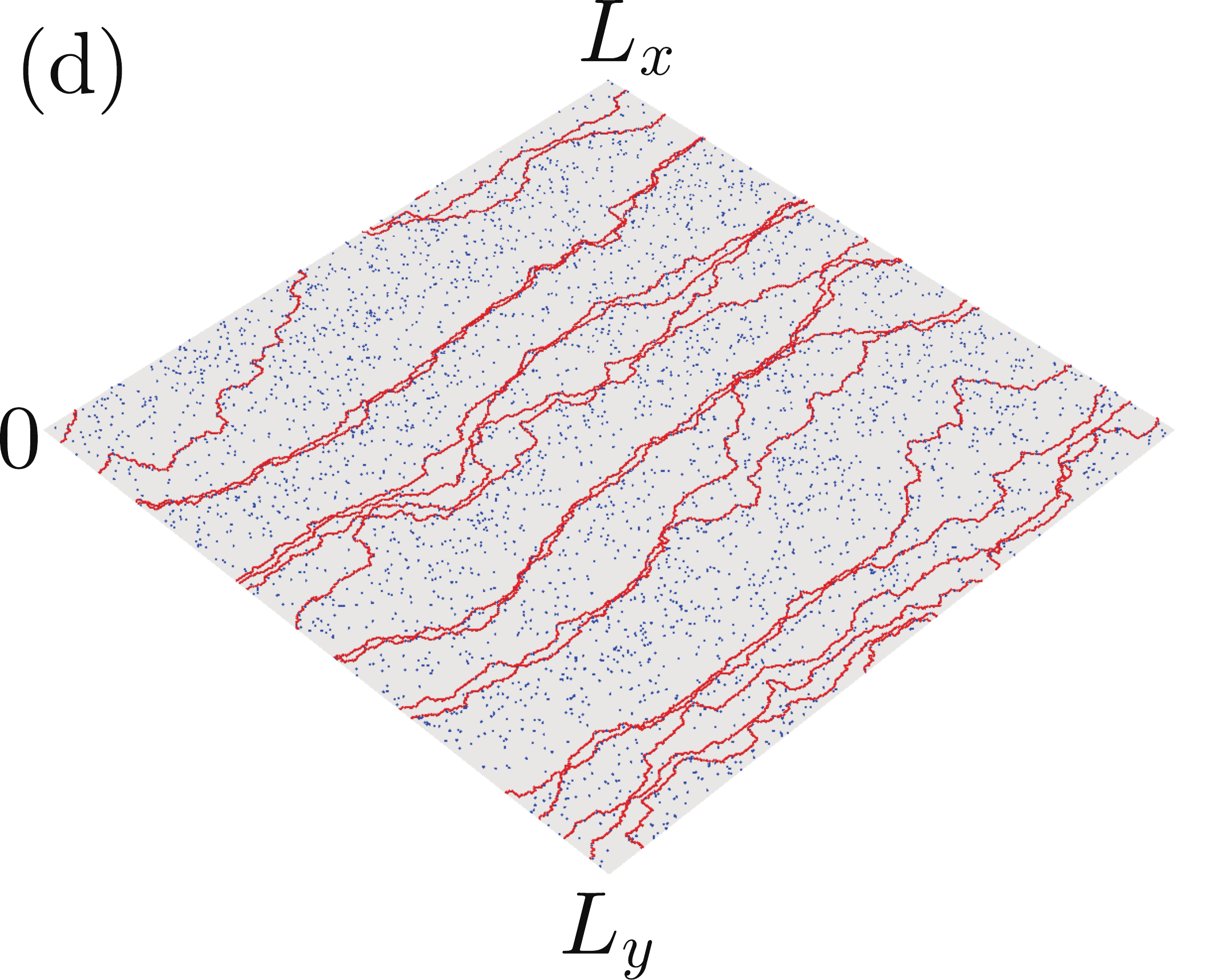} 
\vspace*{0.5cm}

\includegraphics[width=5.5cm,clip]{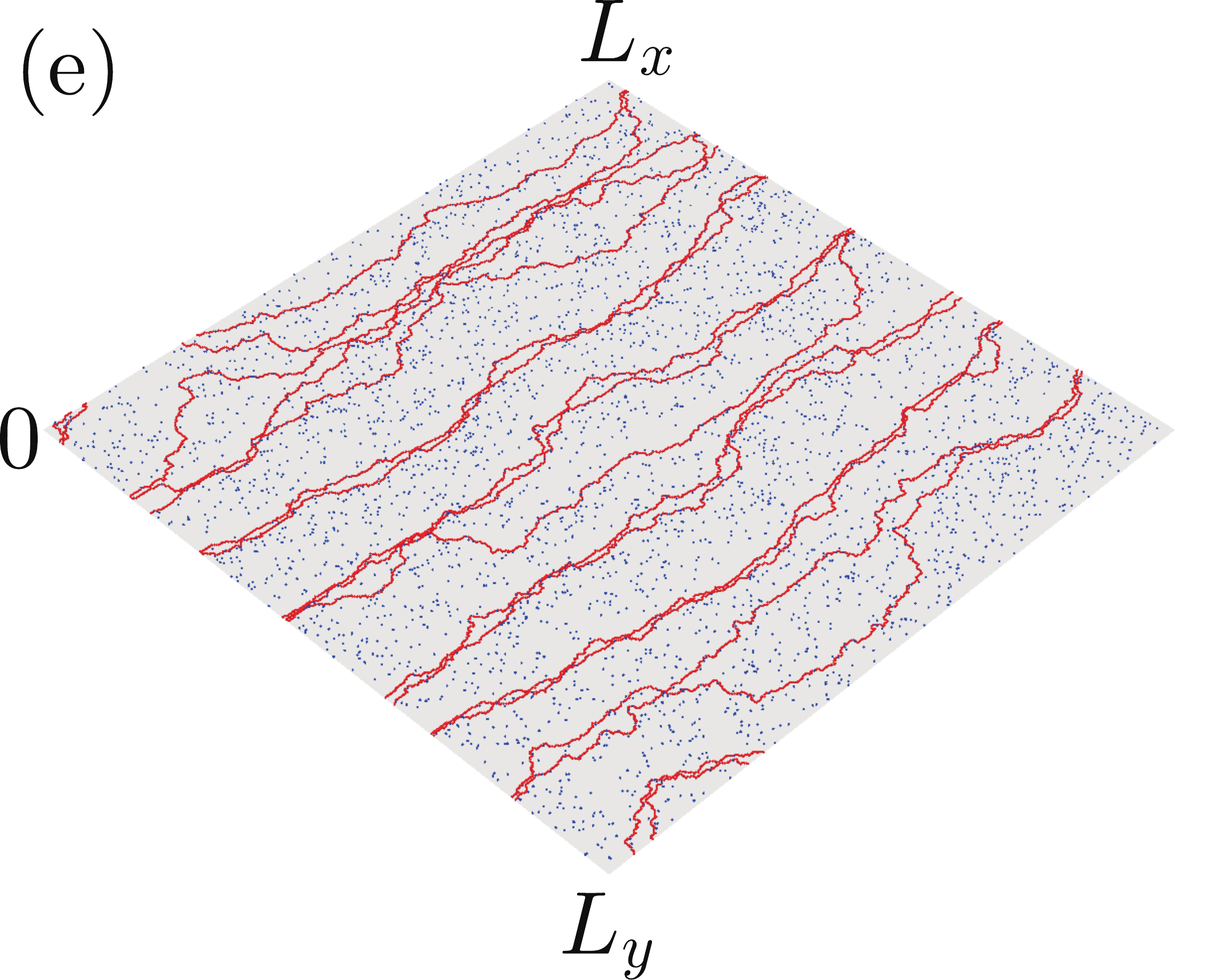} 
\includegraphics[width=5.5cm,clip]{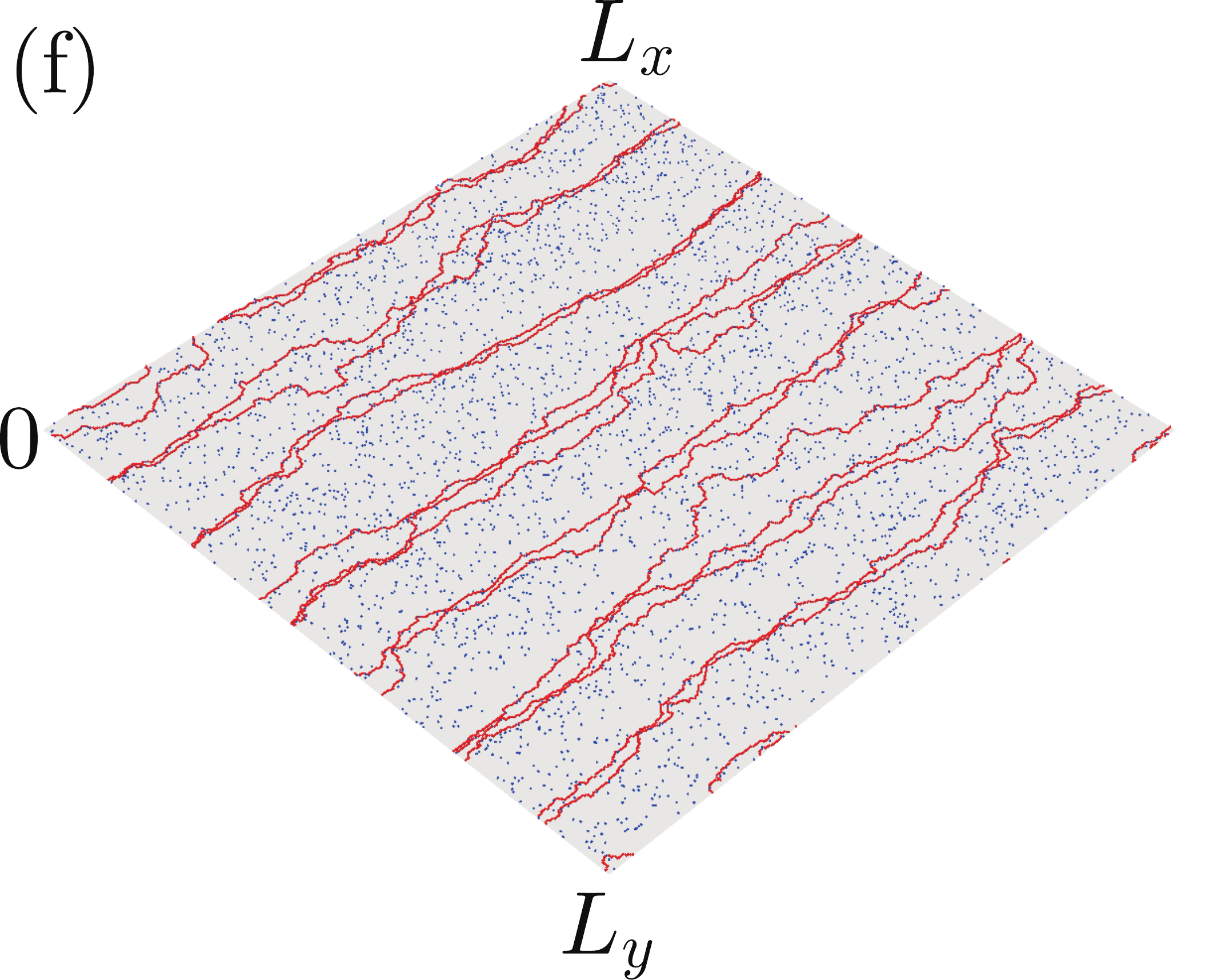}

\caption{
(color online)
Snapshots of a vicinal face  in  late stages during step bunching 
induced by impurities with the surface diffusion of impurities,
where  (a) $D_\mathrm{imp}/D_\mathrm{s}=10^{-5}$
at  $t = 3.5 \times 10^6$,
(b) $D_\mathrm{imp}/D_\mathrm{s}=2 \times 10^{-5}$
at  $t = 3.5 \times 10^6$,
(c) $D_\mathrm{imp}/D_\mathrm{s}=3 \times 10^{-5}$
at  $t = 3.5 \times 10^6$,
(d) $D_\mathrm{imp}/D_\mathrm{s}=4 \times 10^{-5}$
at  $t = 3.6 \times 10^6$,
(e) $D_\mathrm{imp}/D_\mathrm{s}=5 \times 10^{-5}$
at  $t = 3.5 \times 10^6$,
and 
(f) $D_\mathrm{imp}/D_\mathrm{s}=6\times 10^{-5}$
at  $t = 3.6 \times 10^6$.
The blue dots on the surface represent impurities.}
\label{fig:snapshot-migration}
\end{figure}
We set the diffusion of adatoms to be much faster than that of impurities 
and performed  simulations.
Figure~\ref{fig:snapshot-migration} shows 
snapshots of the step bunches formed by impurities.
The ratio  $D_\mathrm{imp}/D_\mathrm{s} $ is
changed from $10^{-5}$  to $6 \times 10^{-5}$.
$N_\mathrm{B}$ seems to  decrease with increasing $D_\mathrm{imp}/D_\mathrm{s} $.
From the snapshots, step bunching without the surface diffusion of impurities is not much different from 
that with the surface diffusion of impurities.
However, the difference in the process of  step bunching between these two cases 
is obvious from the time evolution of average step positions 
(Fig.~\ref{fig:time-evolution-migration}).

\begin{figure}[htbp]
\centering

\includegraphics[width=6.0cm,clip]{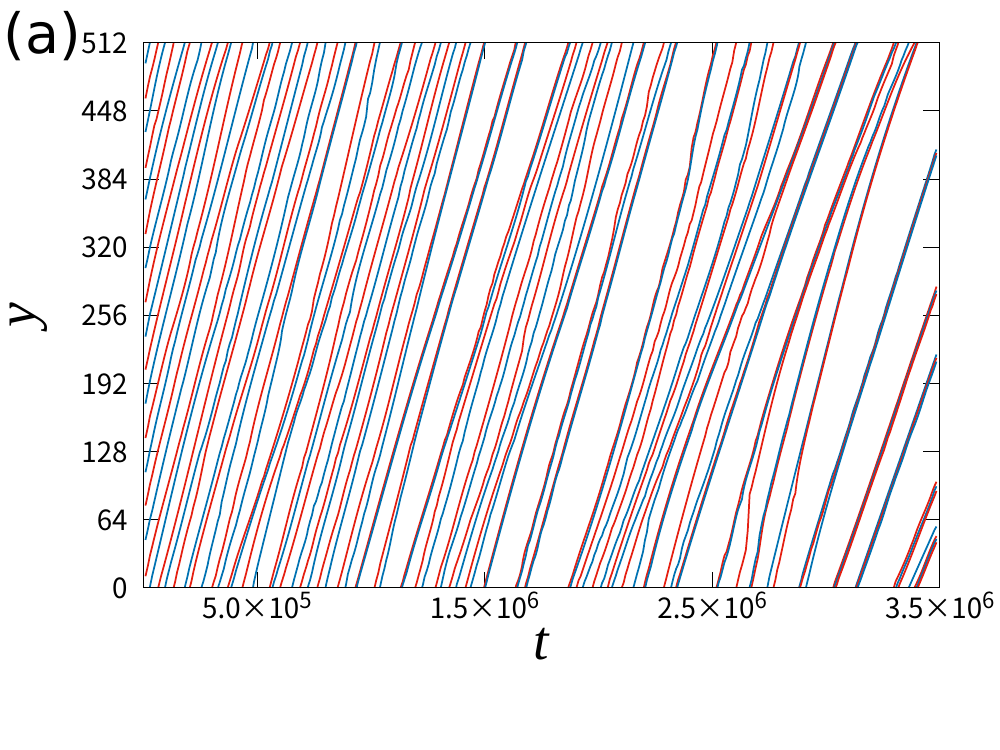} 
\includegraphics[width=6.0cm,clip]{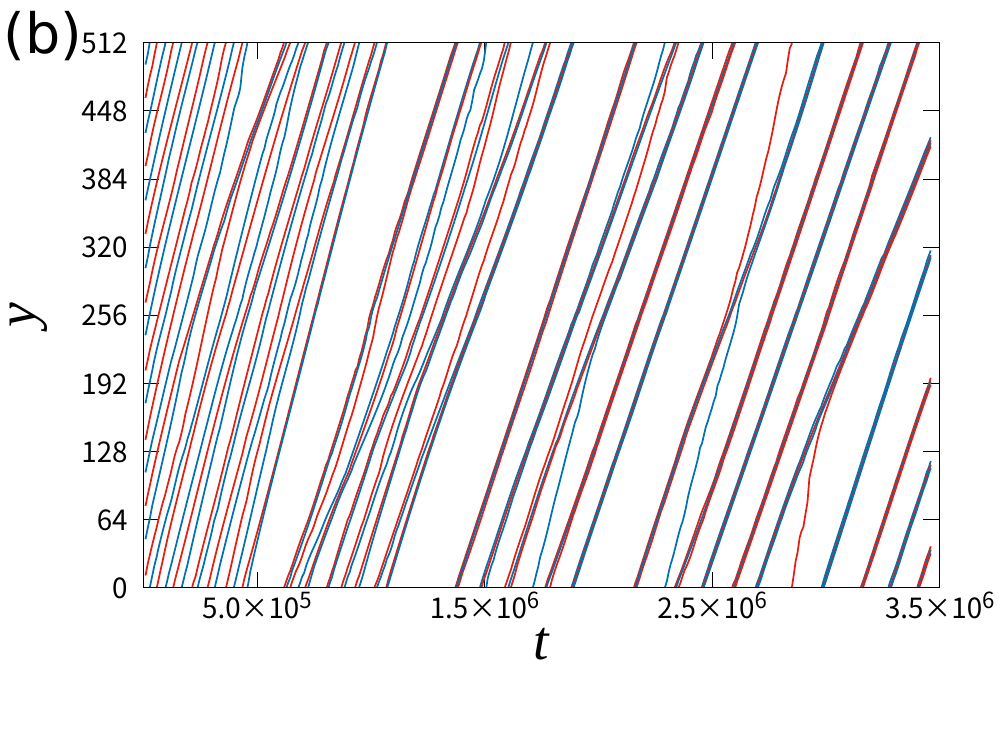} 

\includegraphics[width=6.0cm,clip]{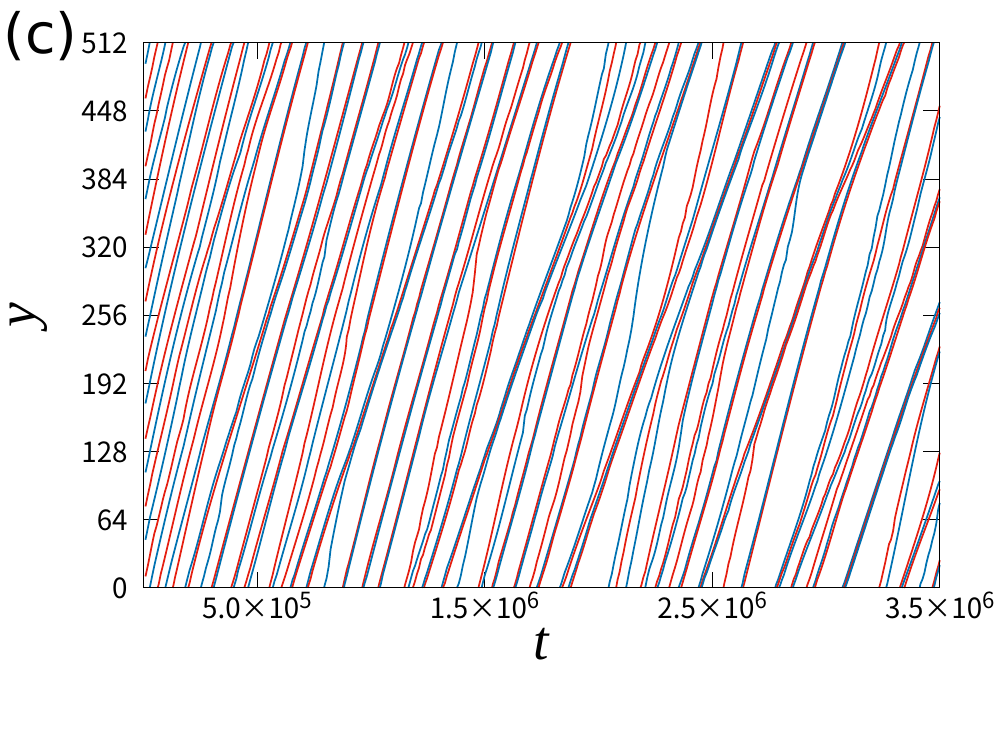} 
\includegraphics[width=6.0cm,clip]{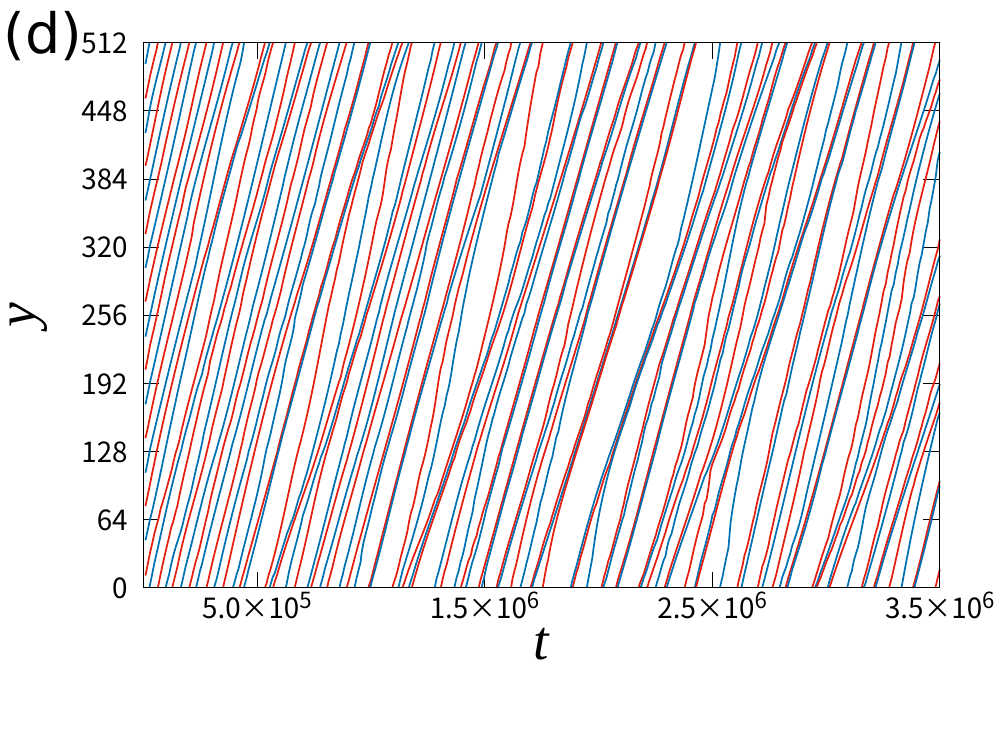} 

\includegraphics[width=6.0cm,clip]{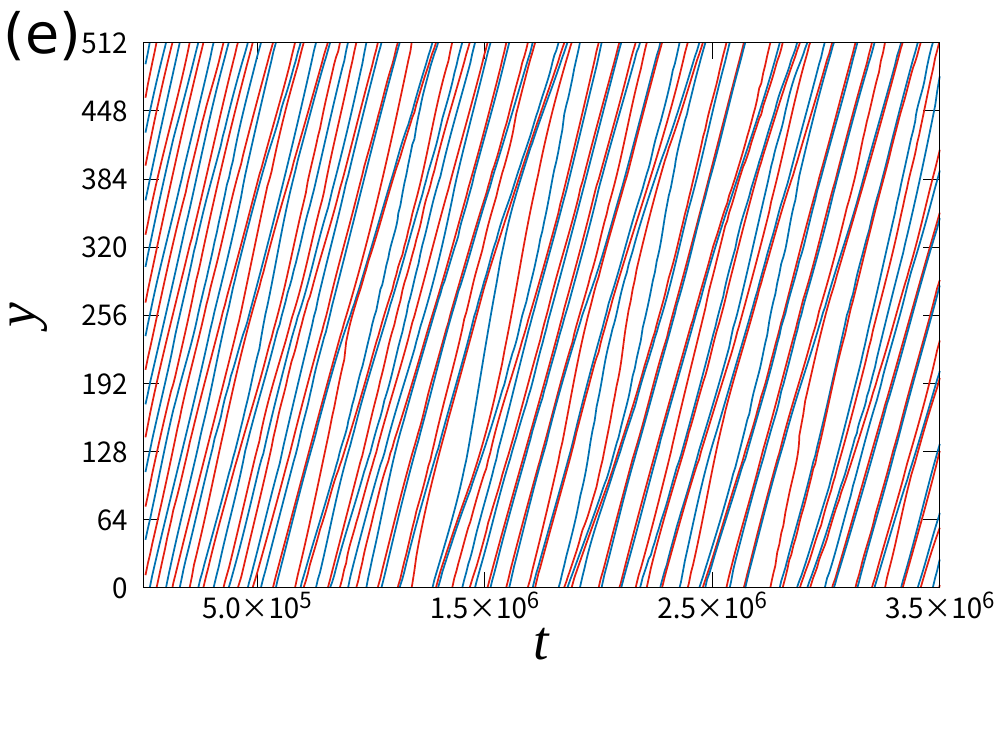} 
\includegraphics[width=6.0cm,clip]{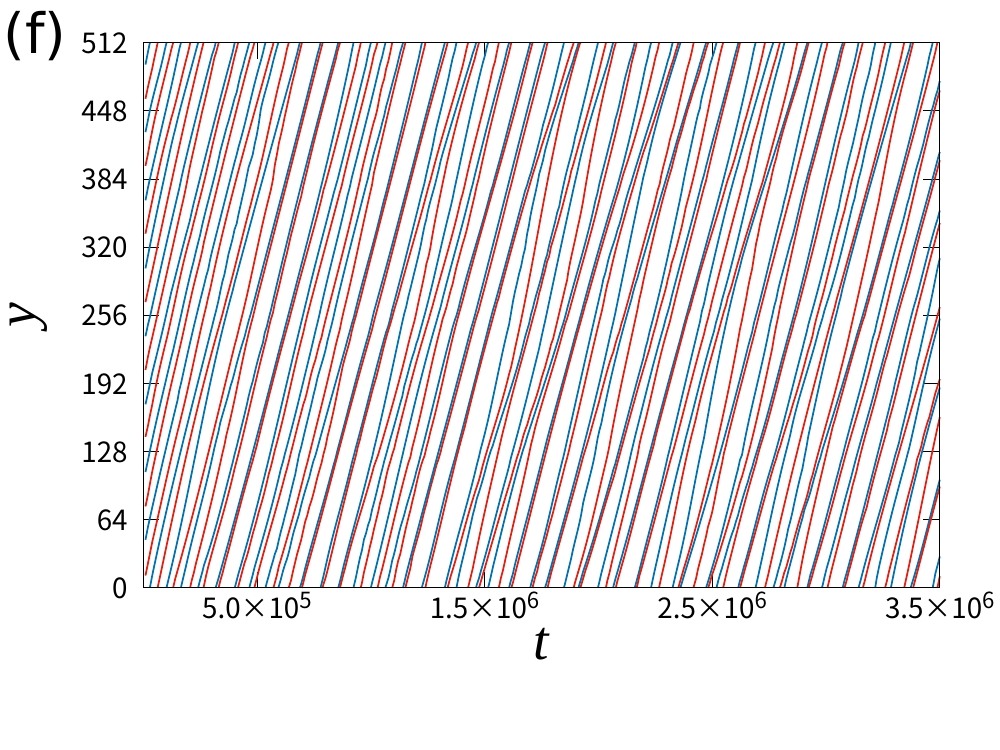} 

\caption{ 
(color online)
Time evolution of step positions averaged in the $x$-direction
for the samples  used in Fig.~\ref{fig:snapshot-migration}(a)--(f).
$\tau_\mathrm{imp}$ is set to $10^5$.
$D_\mathrm{imp}/D_\mathrm{s}$ 
is
(a) $1 \times 10^{-5}$,
(b) $2 \times 10^{-5}$,
(c) $3 \times 10^{-5}$,
(d) $4 \times 10^{-5}$,
(e) $5 \times 10^{-5}$,
and (f) $6 \times 10^{-5}$.
}
\label{fig:time-evolution-migration}
\end{figure}
Figures~\ref{fig:time-evolution-migration}(a)--(f)  show  the time evolution  of the average 
step positions for the samples in obtaining Figs.~\ref{fig:snapshot-migration}(a)--(f), respectively.
Small bunches are formed in an early stage in Figs.~\ref{fig:time-evolution-migration}(a) and (b).
The bunches gather and large bunches separated  by large terraces are formed
in the last stage. 
The process of forming large bunches seems to be  almost the same as that
observed  in Fig.~\ref{fig:timeevolution_tau80000}.
However, 
the separation  and  collision of  single steps
occur in a  later stage in Figs.~\ref{fig:time-evolution-migration}(c)--(f).
In particular,
large bunches are not formed in Figs.~\ref{fig:time-evolution-migration}  (e) and (f). 
The separation and collision between small bunches and single steps are 
frequently repeated.
\begin{figure}[htbp]
\centering

\includegraphics[width=8.0cm,clip]{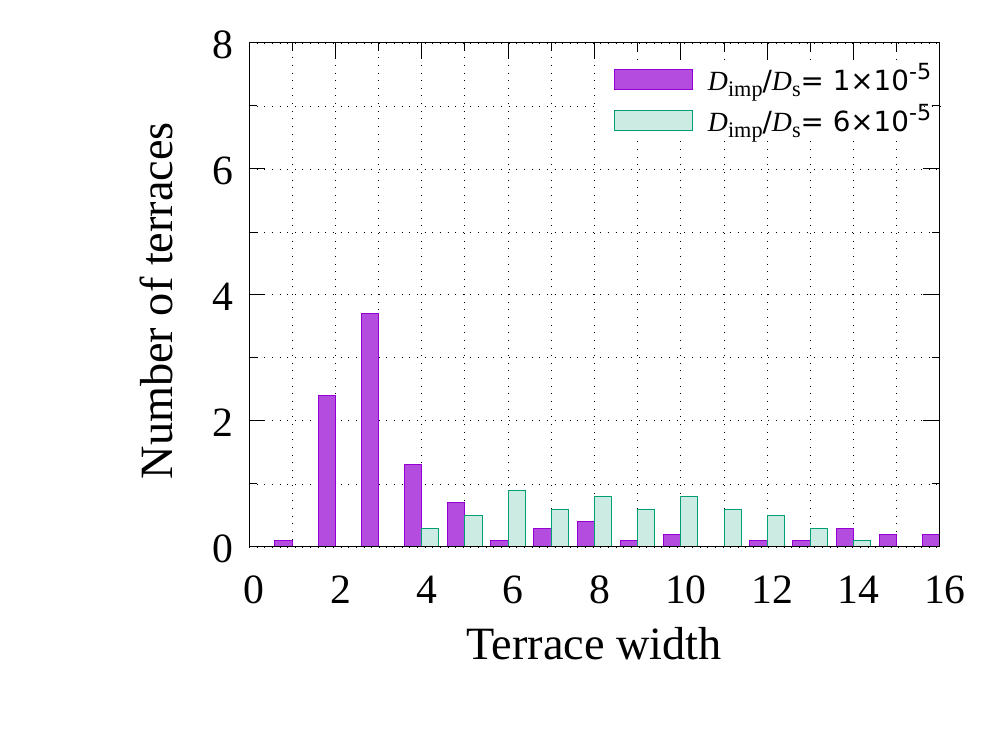} 

\caption{
(color online)
Distributions of terrace widths  that  are smaller than 16
for $D_\mathrm{imp}/D_\mathrm{s}= 1 \times 10^{-5}$ and 
$D_\mathrm{imp}/D_\mathrm{s}= 6 \times 10^{-5}$.
The data are averaged over 10 individual runs.
}
\label{fig:terrace_width_diff}
\end{figure}
They  affect the distribution of terrace widths.
Because of the separation and collision of steps, not only the average terrace width becomes large 
but also the distribution of terrace widths  becomes broad 
(Fig.~\ref{fig:terrace_width_diff}).

\begin{figure}[htbp]
\centering

\includegraphics[width=8.0cm,clip]{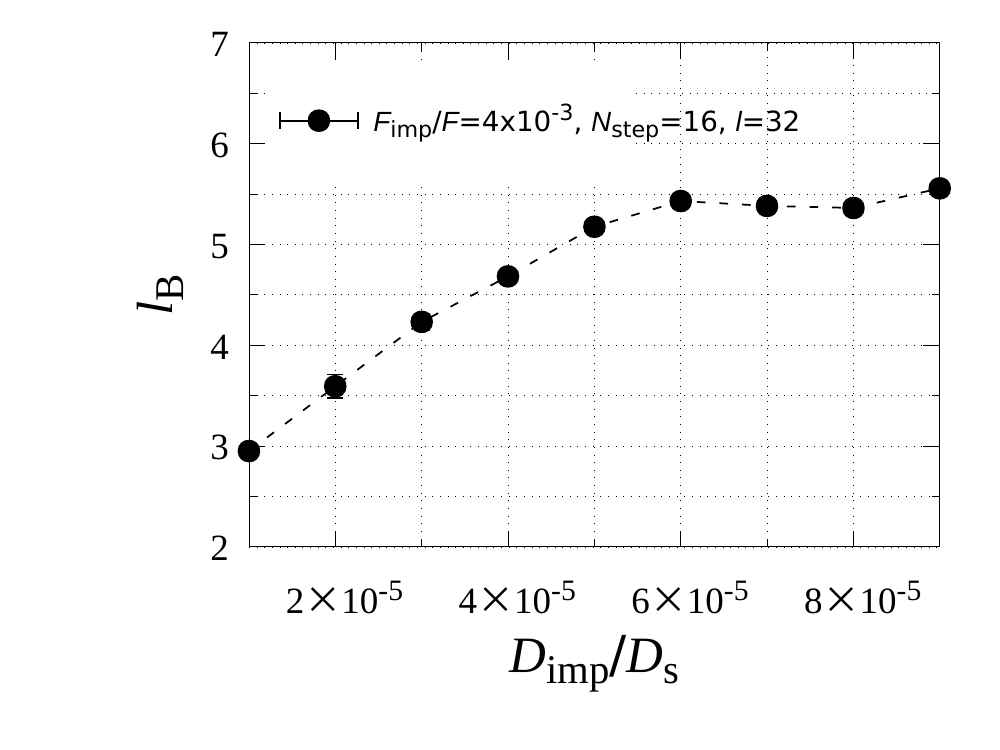} 

\caption{
Dependence of $l_\mathrm{B}$   on $D_\mathrm{imp}/D_\mathrm{s}$.
The data are averaged over 10 individual runs.
}
\label{fig:ave-l_diff}
\end{figure}
Figure~\ref{fig:ave-l_diff} shows the dependence of $l_\mathrm{B}$  on $D_\mathrm{imp}/D_\mathrm{s}$,
where $l_c$ is the same as that used in Fig.~\ref{fig:number_singlesteps}.
$l_\mathrm{B}$  increases with  increasing $D_\mathrm{imp}/D_\mathrm{s}$
and seems to saturate in a large  $D_\mathrm{imp}/D_\mathrm{s}$ region,
where the separation and collision of steps frequently occur.
The saturated value of $l_\mathrm{B}$ is larger than the maximum value of $l_\mathrm{B}$  in Fig.~\ref{fig:ave-l}.

\begin{figure}[htbp]
\centering

\includegraphics[width=8.0cm,clip]{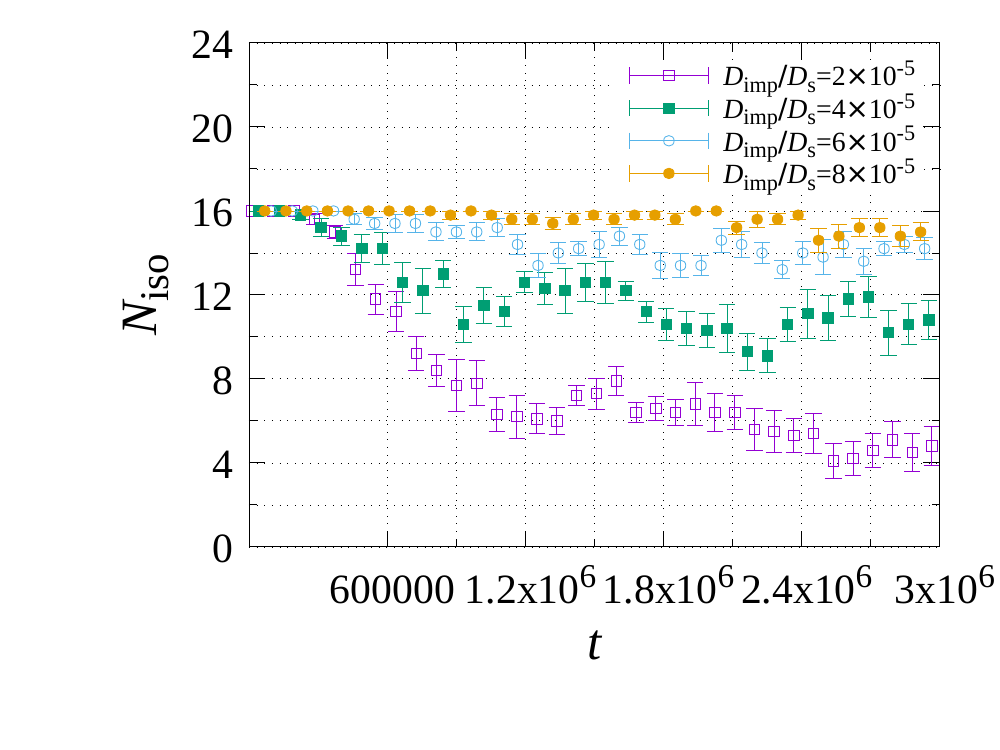} 

\caption{
(color online)
Time dependence of $N_\mathrm{iso}$  on $D_\mathrm{imp}/D_\mathrm{s}$. 
The data are averaged over 10 individual runs.
}
\label{fig:number_singlesteps_diff}
\end{figure}
\begin{figure}[htbp]
\centering

\includegraphics[width=8.0cm,clip]{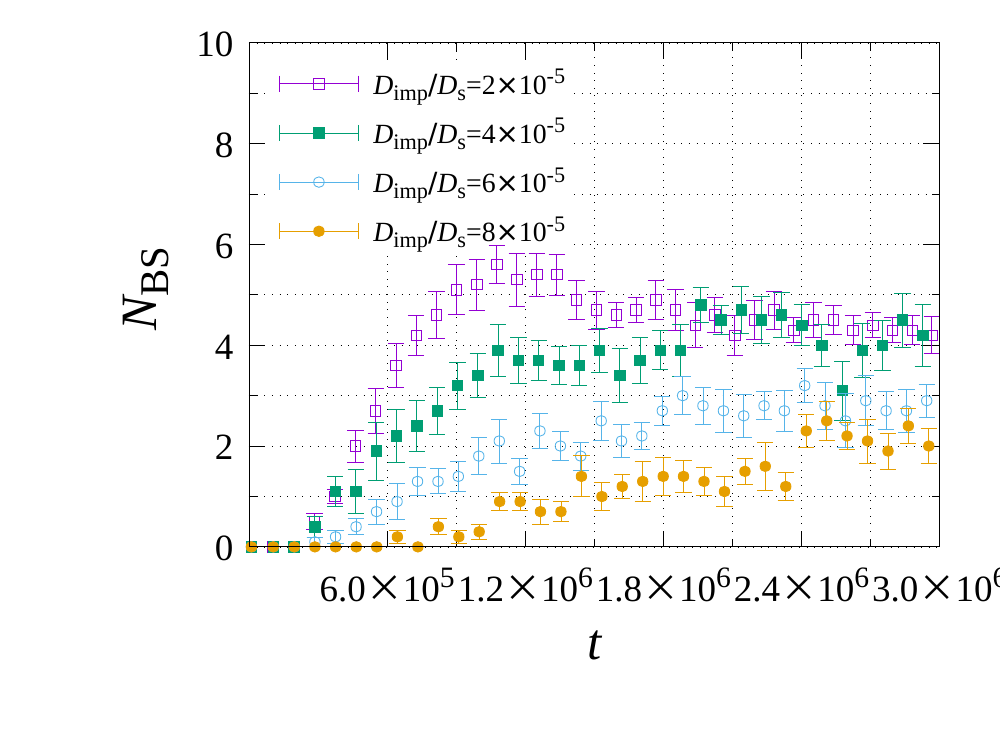} 

\caption{
(color online)
Time dependence of $N_\mathrm{BS}$.
The data are averaged over 10 individual runs.
}
\label{fig:bunch_size_diff}
\end{figure}
\begin{figure}[htbp]
\centering

\includegraphics[width=8.0cm,clip]{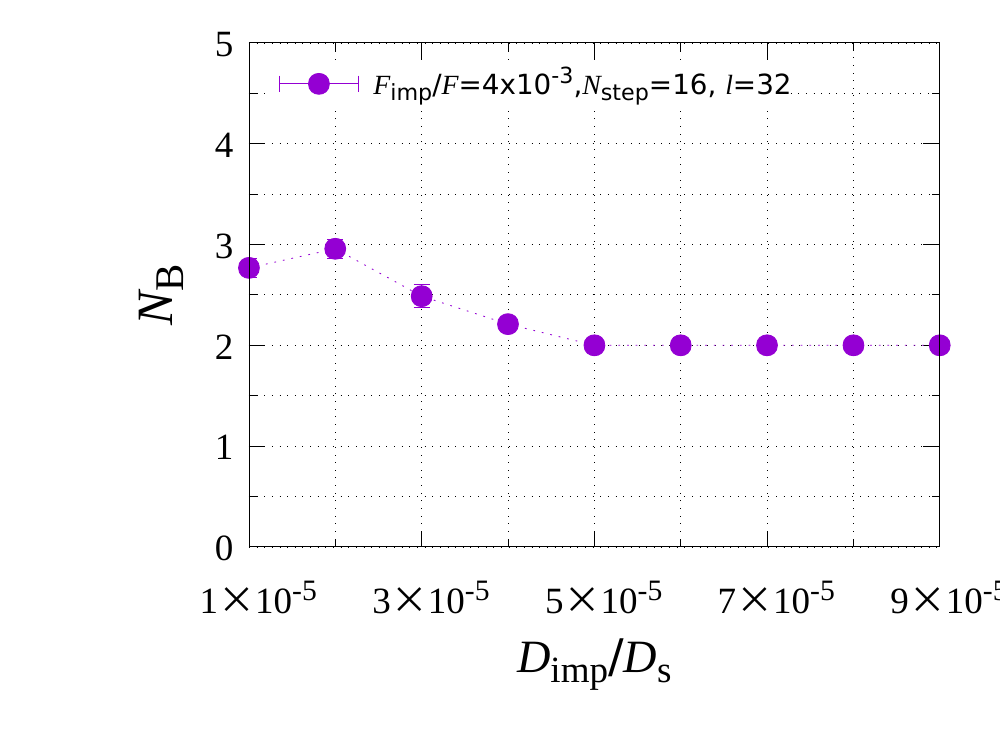} 

\caption{
(color online)
Dependence of $N_\mathrm{B}$ in the last stage on  $D_\mathrm{imp}/D_\mathrm{s}$.
The data are averaged over 10 individual runs.
}
\label{fig:final_bunch_size_diff}
\end{figure}
We analyze the  properties of step bunches in more details.
Figure~\ref{fig:number_singlesteps_diff} shows the time dependence of  $N_\mathrm{iso}$ 
for some $D_\mathrm{imp}/D_\mathrm{s}$.
 $N_\mathrm{iso}$  decreases with increasing time due  to the formation of step bunches
but 
in the last stage increases with increasing $D_\mathrm{imp}/D_\mathrm{s}$.
Almost all the steps are single steps when $D_\mathrm{imp}/D_\mathrm{s}=8 \times 10^{-5}$.
The time evolution of $N_\mathrm{BS}$ is opposite to that of $N_\mathrm{iso}$
(Fig.~\ref{fig:bunch_size_diff}):
$N_\mathrm{BS}$ increases with increasing time and the saturated value in the last stage  
decreases with increasing $D_\mathrm{imp}/D_\mathrm{s}$
(Fig.~\ref{fig:final_bunch_size_diff}), in which 
the final value of $N_\mathrm{B}$  is not more than three. 
From Figs.~\ref{fig:number_singlesteps_diff}--\ref{fig:final_bunch_size_diff}, 
we find that a few  small bunches form and step separation
 frequently occurs  when $D_\mathrm{imp}/D_\mathrm{s}$ is large.

\begin{figure}[htbp]
\centering

\includegraphics[width=8.0cm,clip]{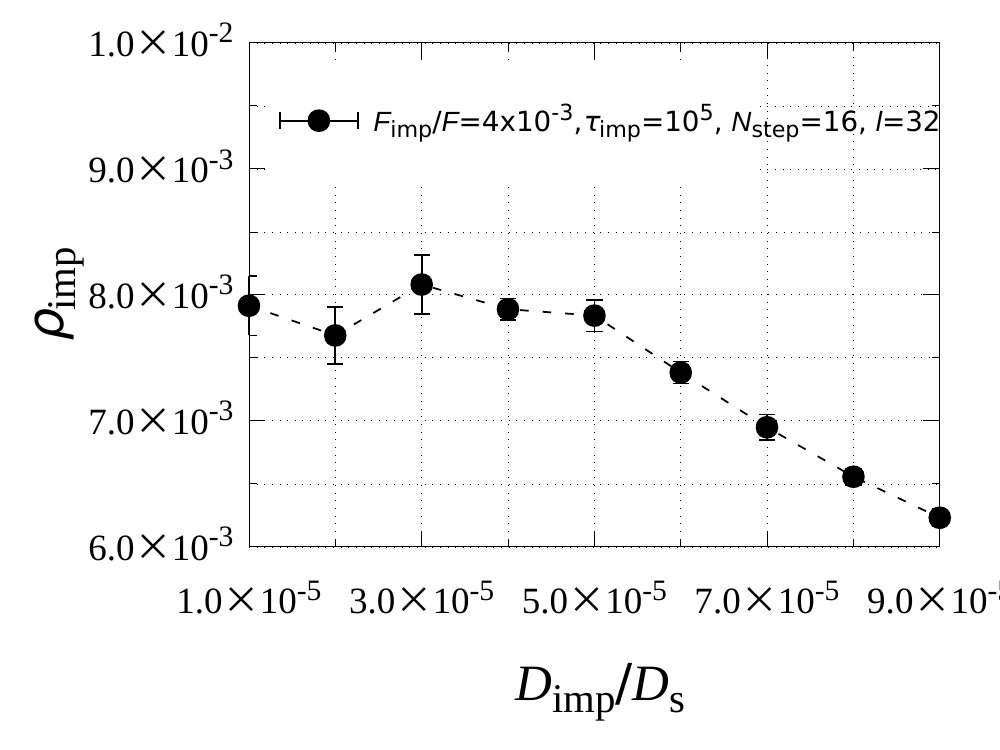} 

\caption{
Dependence of $\rho_\mathrm{imp}$ in the last stage
on $D_\mathrm{imp}/D_\mathrm{s}$.
The data are averaged over 10 individual runs.
}
\label{fig:impurity_insol_migration}
\end{figure}
\begin{figure}[htbp]
\centering

\includegraphics[width=8.0cm,clip]{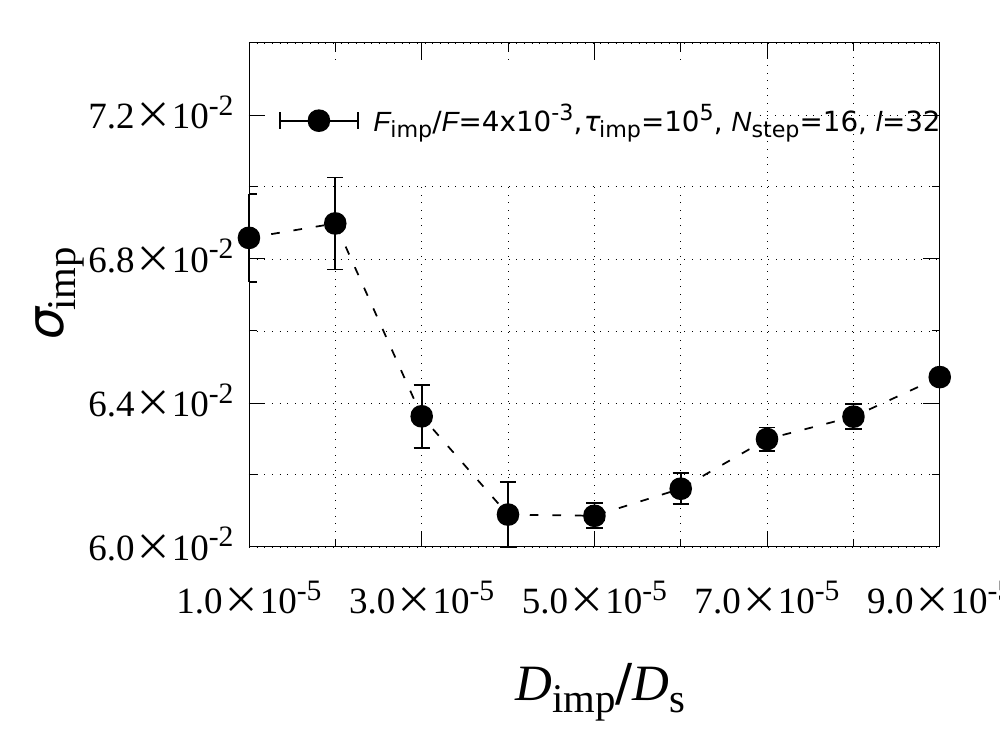} 

\caption{
Dependence of $\sigma_\mathrm{imp}$ in the last stage
on $D_\mathrm{imp}/D_\mathrm{s}$.
The data are averaged over 10 individual runs.
}
\label{fig:impurity_onsurface_migration}
\end{figure}
\begin{figure}[htbp]
\centering

\includegraphics[width=8.0cm,clip]{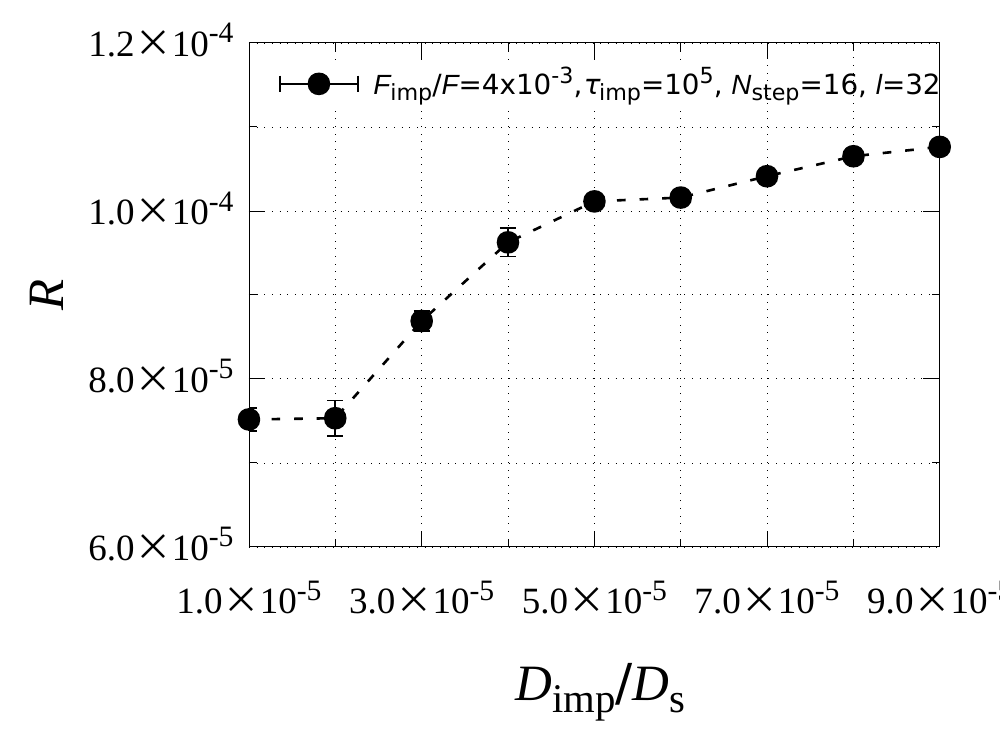} 

\caption{
Dependence of  $R$ on $D_\mathrm{imp}/D_\mathrm{s}$.
The data are averaged over 10 individual runs.
}
\label{fig:r_migration}
\end{figure}
We show how $\sigma_\mathrm{imp}$,  $\rho_\mathrm{imp}$, 
and the growth rate of a vicinal face, $R$, depend on  $D_\mathrm{imp}/D_\mathrm{s}$
in Figs.~\ref{fig:impurity_insol_migration}, \ref{fig:impurity_onsurface_migration}, and \ref{fig:r_migration}, respectively.
From Figs.~\ref{fig:time-evolution-migration}(a) and (b),
we find that the separation and  collision of  single steps hardly occur
when $D_\mathrm{imp}/D_\mathrm{s}=1\times 10^{-5}$ and $2 \times 10^{-5}$.
The surface diffusion of impurities is slow in these cases,
so that the impurities in front of  the step  bunches prevent the lowest steps in the step  bunches 
from advancing faster than the other steps in the bunches.
Therefore the separation of steps from bunches does not occur,
and $\rho_\mathrm{imp}$ is large because almost all the impurities in front of the step bunches are incorporated into the  solid 
when the step bunches advance.
When $D_\mathrm{imp}/D_\mathrm{s}$ is large,  the impurities in front of the step bunches
move away though  surface diffusion before being incorporated into the  solid.
The effect that the impurities prevent the lowest steps from advancing is weakened.  
The impurities in front of the step bunches also move in the step bunches.
These  impurities prevent the steps in bunches from catching up to the lowest steps.
Hence, separation of steps from step bunches becomes possible. 
When this separation,
$R$ increases as  separated steps move faster than step bunches,
and $\sigma_\mathrm{imp}$ decreases because the impurities swept by the separated steps increases. 
However, when $D_\mathrm{imp}/D_\mathrm{s}$ increases further, 
so many impurities move away before being incorporated into solid at single steps.
Therefore,  $\sigma_\mathrm{imp}$ increases again and $\rho_\mathrm{imp}$ decreases.

\section{Summary}\label{sec:summary}
We performed  Monte Carlo simulations and 
studied the effects of both the  evaporation  and  diffusion 
of impurities on  step bunching  induced by impurities.
When we take into account  the evaporation of impurities
and neglect the surface diffusion of impurities, the step bunching proceeds by the collision of small bunches.
When the evaporation of impurities increases, 
the effect of impurities on forming bunches is weakened and the size of bunches 
decreases,
but the separation of steps from bunches does not occur in this instance.

The surface diffusion of impurities also weakens the effect of impurities  on the formation of large bunches,
and the separation of single steps  occurs.
The front of a bunch is the dirtiest area with impurities
because the area is exposed to the vapor phase for the  longest time.
If the lowest step  in a bunch  tries  to escape from the step bunches in the system 
without surface diffusion  of impurities,  
the second lowest step in the bunch  easily catches up with the lowest step
because there are few impurities in front of the second lowest step.
However, impurities can move even in the step bunch when the surface diffusion of impurities occurs.
The impurities coming in front of the second lowest step 
probably prevent the second lowest step from catching up with the lowest step.
Therefore,
 the lowest step can separate from the step bunch if the surface diffusion of impurities is sufficiently fast.
We do not have simulation results showing  evidence for the above scenario directly.
However, when we take into  account  Figs.~\ref{fig:impurity_insol_migration}--\ref{fig:r_migration},
we believe  the scenario we mentioned above is  reasonable in explaining  why 
the separation of steps occurs in the system with the surface diffusion of impurities.

\begin{acknowledgments}
This work is supported by JSPS KAKENHI Grant Numbers JP16K05470, JP18H03839,
18K04960, and the Grant for Joint Research Program of the Institute 
of Low Temperature Science, Hokkaido University,  Grant Number 19G020.
\end{acknowledgments}

\end{document}